\documentclass[twocolumn,showpacs,showkeys,preprintnumbers,amsmath,amssymb,prd,letterpaper,floatfix,superscriptaddress,10pt,aps,nofootinbib]{revtex4-2}
\usepackage[utf8]{inputenc}
\usepackage{textcomp}
\usepackage{graphicx}
\usepackage{color}
\usepackage{longtable}
\usepackage[breaklinks=true]{hyperref}
\usepackage{latexsym}
\usepackage{amsmath}
\usepackage{amssymb}
\usepackage{dsfont}
\usepackage{verbatim}
\usepackage{bm}
\usepackage{bbm}
\usepackage{placeins}
\usepackage{newunicodechar}

\inputencoding{utf8}
\newunicodechar{γ}{$\gamma$}
\newunicodechar{ρ}{$\rho$}
\newunicodechar{π}{$\pi$}
\newunicodechar{Δ}{$\Delta$}
\newunicodechar{η}{$\eta$}
\newunicodechar{ψ}{$\psi$}

\hypersetup{colorlinks, linkcolor = [rgb]{0,0.0,0.75}, citecolor = [rgb]{0,0.0,0.75}, urlcolor = [rgb]{0,0.0,0.75}}

\newcommand{\I}{\mathrm{i}}

\newcommand{\epow}[1]{\mathrm{e}^{ #1 }}
\newcommand{\Tr}[1]{\mathrm{Tr}\big( #1 \big)}

\newcommand{\dbar}{\bar{d}}
\newcommand{\qbar}{\bar{q}}

\newcommand{\pvec}{\vec{p}}
\newcommand{\xvec}{\vec{x}}

\newcommand{\qvec}{\vec{q}}
\newcommand{\yvec}{\vec{y}}
\newcommand{\zvec}{\vec{z}}

\newcommand{\vecQ}{\vec{q}}
\newcommand{\vecP}{\vec{P}}
\newcommand{\vecp}{\vec{p}_{\pi}}

\newcommand{\tSrc}{t_{\pi\pi}}
\newcommand{\tSnk}{t_{\pi}}
\newcommand{\tCurr}{t_{J}}
\newcommand{\DeltaT}{\Delta t}
\newcommand{\ampV}{{\cal V}_{\pi\gamma \to \pi\pi}}

\newcommand{\formF}{F}
\newcommand{\ampT}{{\cal T}_{\pi\pi \to \pi\pi}}

\def\MIT{Center for Theoretical Physics, Laboratory for Nuclear Science and Department of Physics, Massachusetts Institute of Technology, Cambridge, MA 02139, USA}
\def\RFWUB{Helmholtz-Institut f\"ur Strahlen- und Kernphysik, Rheinische Friedrich-Wilhelms-Universit\"at Bonn, Nu{\ss}allee 14-16, D-53115 Bonn, Germany}
\def\CYI{Computation-based Science and Technology Research Center, Cyprus Institute, 20 Kavafi Str., 2121 Nicosia, Cyprus}
\def\UCY{Department of Physics, University of Cyprus, P.O. Box 20537, 1678 Nicosia, Cyprus}
\def\RIKEN{RIKEN BNL Research Center, Brookhaven National Laboratory, Upton, NY 11973, USA}

\def\UOA{Department of Physics, University of Arizona, Tucson, AZ 85721, USA}
\def\SBU{Department of Physics and Astronomy, Stony Brook University, Stony Brook, NY 11794, USA}

\begin{document}
%%%%%%%%%%%%%%%%%%%%%%%%%%%%%%%%%%%%%%%%%%%%%%%%%%%%%%%%%

\author{Constantia Alexandrou}
 \affiliation{\UCY}
 \affiliation{\CYI}

\author{Luka Leskovec}
\email{leskovec@email.arizona.edu}
 \affiliation{\UOA}

\author{Stefan Meinel}
\email{smeinel@email.arizona.edu}
 \affiliation{\UOA}
 \affiliation{\RIKEN}

\author{John Negele}
 \affiliation{\MIT}

\author{\mbox{Srijit Paul}}
 \affiliation{\CYI}

\author{Marcus Petschlies}
\email{marcus.petschlies@hiskp.uni-bonn.de}
 \affiliation{\RFWUB}

\author{Andrew Pochinsky}
 \affiliation{\MIT}

\author{Gumaro Rendon}
 \affiliation{\UOA}

\author{Sergey Syritsyn}
 \affiliation{\RIKEN}
 \affiliation{\SBU}

\date{December 27, 2021}
%%%%%%%%%%%%%%%%%%%%%%%%%%%%%%%%%%%%%%%%%%%%%%%%%%%%%%%%%%%%%%%%%%%%%%%%%%%%%%%%%%%%%%%%%%%%%%%%%%%%%%%%%%%%%%%
\title{\texorpdfstring{The $\bm{\pi\gamma \to \pi \pi}$ transition and the $\bm{\rho}$ radiative decay width from lattice QCD}{}}
%%%%%%%%%%%%%%%%%%%%%%%%%%%%%%%%%%%%%%%%%%%%%%%%%%%%%%%%%%%%%%%%%%%%%%%%%%%%%%%%%%%%%%%%%%%%%%%%%%%%%%%%%%%%%%%

\begin{abstract}
We report a lattice QCD determination of the $\pi\gamma \to \pi \pi$ transition amplitude for the $P$-wave, $I=1$ two-pion final state,
as a function of the photon virtuality and $\pi\pi$ invariant mass. The calculation was performed
with $2+1$ flavors of clover fermions at a pion mass of approximately $320$ MeV, on a $32^3\times96$ lattice with $L\approx 3.6$ fm.
We construct the necessary correlation functions using a combination of smeared forward, sequential and stochastic propagators,
and determine the finite-volume matrix elements for all $\pi\pi$ momenta up to $|\vec{P}|= \sqrt{3} \frac{2\pi}{L}$ and
all associated irreducible representations. In the mapping of the finite-volume to infinite-volume matrix elements using the
Lellouch-L\"uscher factor, we consider two different parametrizations of the $\pi\pi$ scattering phase shift. We fit the $q^2$
and $s$ dependence of the infinite-volume transition amplitude in a model-independent way using series expansions, and compare
multiple different truncations of this series. Through analytic continuation to the $\rho$ resonance pole, we
also determine the $\pi\gamma \to \rho$ resonant transition form factor and the $\rho$ meson photocoupling, and obtain $|G_{\rho\pi\gamma}| = 0.0802(32)(20)$.
\end{abstract}

\maketitle

%%%%%%%%%%%%%%%%%%%%%%%%%%%%%%%%%%%%%%%%%%%%%%%%%%%%%%%%%%%%%%%%%%%%%%%%%%%%%%%%%%%%%%%%%%%%%%%%%%%%%%%%%%%%%%%
%%%                                          INTRODUCTION                                                   %%%
%%%%%%%%%%%%%%%%%%%%%%%%%%%%%%%%%%%%%%%%%%%%%%%%%%%%%%%%%%%%%%%%%%%%%%%%%%%%%%%%%%%%%%%%%%%%%%%%%%%%%%%%%%%%%%%

%%%%%%%%%%%%%%%%%%%%%%%%%%%%%%%%%%%%%%%%%%%%%%%%%%%%%%%%%%%%%%%%%%%%%%%%%%%%%%%%%%%%%%%%%%%%%%%%%%%%%%%%%%%%%%%
\section{INTRODUCTION}\label{sec_introduction}
%%%%%%%%%%%%%%%%%%%%%%%%%%%%%%%%%%%%%%%%%%%%%%%%%%%%%%%%%%%%%%%%%%%%%%%%%%%%%%%%%%%%%%%%%%%%%%%%%%%%%%%%%%%%%%%

During the last decade, there has been tremendous progress with lattice QCD calculations of low-energy hadron-hadron scattering
amplitudes and the associated resonances \cite{Briceno:2017max}. While the $S$-matrix is not directly accessible from the lattice,
the L\"uscher quantization condition and its generalizations \cite{Luscher:1990ux,Feng:2004ua,Rummukainen:1995vs,He:2005ey,Kim:2005gf,Christ:2005gi, Hansen:2012tf,Leskovec:2012gb,Gockeler:2012yj,Guo:2012hv,Briceno:2014oea,Briceno:2017tce,Lee:2017igf} relate the infinite-volume
scattering amplitudes (and their resonance poles) with the discrete finite-volume energy spectra computed on the lattice. A widely studied example is $\pi\pi$ scattering in the $P$ wave, $I=1$ channel, where the $\rho$ resonance resides \cite{Aoki:2007rd,Frison:2010ws,Feng:2010es,Lang:2011mn,Aoki:2011yj,Pelissier:2012pi,Dudek:2012xn,Wilson:2015dqa,Bali:2015gji,Bulava:2016mks,Hu:2016shf,Guo:2016zos,Fu:2016itp,Alexandrou:2017mpi}.

Going beyond spectroscopy, Lellouch and L\"uscher also found the relation between finite-volume and infinite-volume $1\to 2$ transition matrix elements
for the case of the nonleptonic weak decay $K \to \pi\pi$ \cite{Lellouch:2000pv}. The formalism was later extended to include all elastic states below the inelastic threshold \cite{Lin:2001ek} and to moving frames \cite{Christ:2005gi}, and more recently to multiple coupled two-body channels \cite{Hansen:2012tf}, matrix elements
of arbitrary external currents with four-momentum transfer \cite{Briceno:2014uqa,Briceno:2015csa}, and $2\to2$ matrix elements \cite{Briceno:2015tza} (see also Refs.~\cite{Meyer:2011um, Bernard:2012bi, Christ:2015pwa,Agadjanov:2014kha,Detmold:2014fpa, Agadjanov:2016fbd} for related work).

The first numerical calculations involving the Lellouch-L\"uscher formalism were performed for $K\to\pi\pi$, providing
an ab-initio Standard-Model prediction of direct CP violation in this process \cite{Blum:2011ng, Ishizuka:2015oja, Bai:2015nea}.
More recently, the generalization of the formalism by Brice\~no, Hansen, and Walker-Loud (BHWL) \cite{Briceno:2014uqa} was applied by the Hadron Spectrum Collaboration to compute the $\pi\gamma \to \pi\pi$ amplitude, with the $\pi\pi$ system in a $P$-wave, as a function of photon virtuality and $\pi\pi$ invariant mass \cite{Briceno:2015dca, Briceno:2016kkp}. This amplitude describes $\rho$ photoproduction and radiative decay \cite{ODonnell:1981hgt,Landsberg:1986fd}, and also plays an important role in dispersion relations used to calculate the hadronic contributions to the anomalous magnetic moment of the muon \cite{Hoferichter:2012pm,Hoferichter:2014vra,Colangelo:2014pva,Gerardin:2017ryf}. Various theoretical aspects of the $\pi\gamma \to \pi\pi$
process have also been discussed in Refs.~\cite{Terentev:1971cso, Bijnens:1989ff, Ametller:2001yk, Truong:2001en, Hannah:2001ee, Kaiser:2008ss, Hoferichter:2017ftn}.
As far as the finite-volume formalism is concerned, the $\pi\gamma \to \pi\pi$ amplitude in the $\rho$ resonance region is one of the simplest $1\to 2$ processes to study on the lattice, because the $\pi\pi$ scattering is almost completely elastic in the relevant energy region.

In this paper, we report a lattice QCD calculation of the $\pi\gamma \to \pi\pi$ transition with $2+1$ flavors of clover-improved Wilson fermions \cite{Sheikholeslami:1985ij} at a pion mass of approximately $320$ MeV, building upon our previous work on $\pi\pi$ scattering \cite{Alexandrou:2017mpi}.
In contrast to the original Lellouch-L\"uscher approach to the nonleptonic $K\to\pi\pi$ decay, where the lattice parameters need to be tuned such that the
final and initial hadronic states have equal energy, the BHWL formalism enables us to obtain the $\pi\gamma \to \pi\pi$ amplitude for all $\pi\pi$ energy levels
and arbitrary momentum transfer.

In Sec.~\ref{sec_about_photproduction}, we discuss the $\pi\gamma \to \pi\pi$ amplitude and related quantities in the continuum. The parameters of our lattice
calculation are given in Sec.~\ref{sec_lattice}. We describe the interpolating fields and correlation functions in Sec.~\ref{sec_interpolators_and_3pt}, and the
extraction of the finite-volume matrix elements from these correlation functions in Sec.~\ref{sec_matrix_elements}. The mapping from finite volume to infinite
volume using the Lellouch-L\"uscher factor is explained in Sec.~\ref{subsec_LLmap}. We carefully study a model-independent approach for parametrizing
the $q^2$ and $s$ dependence of the $\pi\gamma \to \pi\pi$ amplitude in Sec.~\ref{sec_multihadron}, and present our results for the
 $\pi\gamma \to \pi\pi$ cross section, the $\pi\gamma \to \rho$ resonant form factor, the $\rho$ meson photocoupling, and the $\rho$ radiative decay width
in Sec.~\ref{sec_physics}.

%%%%%%%%%%%%%%%%%%%%%%%%%%%%%%%%%%%%%%%%%%%%%%%%%%%%%%%%%%%%%%%%%%%
\section{\texorpdfstring{ABOUT THE $\pi\gamma \to \pi \pi$ process}{}}\label{sec_about_photproduction}
%%%%%%%%%%%%%%%%%%%%%%%%%%%%%%%%%%%%%%%%%%%%%%%%%%%%%%%%%%%%%%%%%%%
The resonance photoproduction process $\pi \gamma \to \rho$ is obtained from the more general process $\pi\gamma \to \pi \pi$, where the final $\pi\pi$ state is in $P$-wave and couples strongly to the $\rho$ resonance with isospin $I=1,I_3=1$ and $J^{PC}=1^{--}$. Throughout this paper (except where stated otherwise), we allow the photon to be virtual, but continue to denote it as just $\gamma$. The $\pi\pi$ photoproduction is described by the continuum infinite-volume matrix element $\langle \pi\pi | J^\mu(0) | \pi \rangle$, which is constructed from the initial state $|\pi\rangle$, the insertion of the QED current $J^\mu$ (defined without the factor of $e$) and the final state $|\pi\pi\rangle$ with $I=1$ and four-momentum $P=(\sqrt{s+\vecP^2},\vecP)$. The latter is projected to the $P$-wave, so that it couples to the $\rho$ resonance, where the polarization of the system is described by $\epsilon^\nu(P,m)$ \cite{Chung:1971ri}. Due to the Lorentz symmetry the matrix element decomposes like
\begin{align}
\label{eq:LLdecomp}
\langle \pi\pi | J^\mu(0) | \pi \rangle = \frac{2\I \ampV(q^2,s)}{m_\pi} \epsilon^{\nu\mu\alpha\beta} \epsilon_{\nu}(P,m) (p_\pi)_\alpha P_\beta,
\end{align}
where $q=p_\pi-P$ is the photon four-momentum transfer. Above, the current is taken in position space, and the single-pion state is normalized as
\begin{align}
 \langle \pi, \vecp^{\:\prime} | \pi, \vecp \rangle &= 2 E_\pi^{\vecp} (2\pi)^3 \delta^3 (\vecp - \vecp^{\:\prime}).
\end{align}
The $P$-wave two-pion states with polarization $m$ are given by
\begin{align}
 &|\pi\pi, \sqrt{s}, \vecP, 1, m \rangle \cr
 &= \frac{1}{\sqrt{4\pi}} \int \mathrm{d} \widehat{\vec{k}}_{\rm cm}\: Y_{1m}^*(\widehat{\vec{k}}_{\rm cm})  |\pi\pi, \sqrt{s}, \vecP, \widehat{\vec{k}}_{\rm cm} \rangle, \label{eq:IVstates}
\end{align}
where $|\pi\pi, \sqrt{s}, \vecP, \widehat{\vec{k}}_{\rm cm} \rangle$ is a two-pion state with total momentum $\vecP$, relative momentum direction unit vector $\widehat{\vec{k}}_{\rm cm}$ in the center-of-momentum frame, and invariant mass $\sqrt{s}$. These states are normalized according to
\begin{align}
 &\langle \pi\pi, \sqrt{s^\prime}, \vecP^\prime, \widehat{\vec{k}^\prime}_{\!\!\rm cm}   |\pi\pi, \sqrt{s}, \vecP, \widehat{\vec{k}}_{\rm cm} \rangle \cr
 &= 2 E_1\,(2\pi)^3\, 2E_2\,(2\pi)^3\, \delta^3(\vec{k}-\vec{k}^\prime) \delta^3(\vecP-\vec{k}-\vecP^\prime+\vec{k}^\prime), \cr
\end{align}
where $E_1$ and $E_2$ are the individual pion energies,
\begin{align}
 E_1 &= \sqrt{m_\pi^2 + \vec{k}^2} ,\\
 E_2 &= \sqrt{m_\pi^2 + (\vecP-\vec{k})^2}.
\end{align}
These normalizations of states imply that the matrix element \eqref{eq:LLdecomp} is dimensionless and that $\ampV$ has units of ${\rm MeV}^{-1}$.
Notice that there is no explicit $\rho$ label in the amplitude; this is because the $\rho$ is not a QCD asymptotic state, but rather a resonance in $P$-wave $\pi\pi$ scattering with $I=1$ associated with the pole in the scattering amplitude $\ampT$ at $s_P\approx m_R^2 - i m_R \Gamma_R$.
The transition amplitude $\ampV$ depends on both the photon four-momentum transfer $q^2$ and the $\pi\pi$ invariant mass $s$. Like $\ampT$,
this amplitude also has a pole at $s=s_P$; the residue at the pole gives the $\rho$ resonance photoproduction form factor.
For $s$ in the vicinity of $s_P$ and at $q^2=0$, the amplitudes $\ampT$ and $\ampV$ behave like \cite{Briceno:2015csa}
\begin{align}
\label{eq:ampT}
&\ampT(s) \sim \frac{G_{\rho\pi\pi}^2}{s_P - s},\\
\label{eq:ampV}
&\ampV(0, s) \sim \frac{G_{\rho\pi\pi} \:G_{\rho\pi\gamma}}{s_P - s},
\end{align}
where $G_{\rho\pi\pi}$ and $G_{\rho\pi\gamma}$ are the couplings of the $\rho$ resonance to $\pi\pi$ and $\pi\gamma$, respectively.

The $\pi\pi$ elastic scattering amplitude is related to the scattering phase shift $\delta(s)$ via
\begin{align}
\label{eq:T_BW_d}
&\ampT(s) = \frac{16 \pi \sqrt{s}}{k} \frac{1}{\cot{\delta(s)} - \I},
\end{align}
where $k$ is the scattering momentum, defined by $\sqrt{s}=2\sqrt{m_\pi^2 + k^2}$.
Near a narrow resonance, the phase shift is well described by parametrizations of the Breit-Wigner type,
\begin{align}
\label{eq:delta_BW}
&\cot{\delta(s)} = \frac{m_R^2 - s}{\sqrt{s}\Gamma(s)},
\end{align}
where multiple different choices can be used for $\Gamma(s)$. Inserting Eq.~(\ref{eq:delta_BW}) into (\ref{eq:T_BW_d}) gives
\begin{align}
\label{eq:T_BW_s}
\ampT(s) = \frac{16 \pi \sqrt{s}}{k} \frac{\sqrt{s}\Gamma(s)}{m_R^2 - s - \I \sqrt{s}\Gamma(s)}.
\end{align}
Motivated by Eqs.~(\ref{eq:ampT}) and (\ref{eq:ampV}), we write the photoproduction amplitude $\ampV(q^2, s)$ as
\begin{align}
\label{eq:V_BW_d}
\ampV(q^2,s) & =  \sqrt{\frac{16 \pi}{k\Gamma(s)}} \frac{\formF(q^2,s) }{\cot{\delta(s) - \I}} \cr
& =   \sqrt{\frac{16 \pi}{k\Gamma(s)}}  \formF(q^2,s)  \sin{\delta(s)}\, e^{\I\delta(s)},
\end{align}
where the form factor $\formF(q^2,s)$ no longer has a pole in $s$, and becomes equal to the photocoupling $G_{\rho\pi\gamma}$ for $s=m_R^2 - \I m_R \Gamma_R$ and $q^2=0$.
More generally, we define the resonant form factor for arbitrary photon virtuality
as
\begin{equation}
 F_{\pi\gamma\to\rho}(q^2) = \formF(q^2,\, m_R^2 - \I m_R \Gamma_R). \label{eq:Fresonant}
\end{equation}
Note that Eq.~(\ref{eq:V_BW_d}) explicitly satisfies Watson's theorem.

In Ref.~\cite{Alexandrou:2017mpi} we found that our $\pi\pi$ scattering amplitude is well described by the {\bf BWI} and {\bf BWII} Breit-Wigner models discussed in Sec.~II of that same reference, so we will continue to utilize the Breit-Wigner formulas throughout this work. The nonresonant backgrounds were found to be consistent with zero and are not included in the $\pi\pi$ scattering amplitude here. For convenience, we repeat the definitions of {\bf BWI} and {\bf BWII} here:
\begin{itemize}
  \item {\bf BW I:}
  \begin{align}
  \label{eq:Gamma_Pwave}
  \Gamma_{I}(s) = \frac{g_{\rho\pi\pi}^2}{6\pi} \frac{k^{3}}{s},
  \end{align}
  where $g_{\rho\pi\pi}$ is the coupling between the $\pi\pi$ scattering channel and the
  $\rho$ resonance in the Breit-Wigner model.

  \item {\bf BW II:}
  \begin{align}
  \label{eq:Gamma_PwaveBW}
  \Gamma_{II}(s) = \frac{g_{\rho\pi\pi}^2}{6\pi} \frac{k^{3}}{s}\: \frac{1 + (k_R r_0)^2}{1 + (k r_0)^2},
  \end{align}
  where $k_R$ is the scattering momentum at $\sqrt{s}=m_R$ and $r_0$ is the radius
  of the centrifugal barrier \cite{VonHippel:1972fg}.
\end{itemize}

We consider two physically observable quantities we can determine from $|\langle \pi\pi|J_\mu(0)|\pi\rangle|$. The first is the $\pi^+\gamma\to\pi^+\pi^0$ cross section as a function of $\pi^+\pi^0$ invariant mass, which in the center-of-momentum frame is given by \cite{Briceno:2015dca}
\begin{align}
\sigma(\pi^+\gamma\to\pi^+\pi^0; s, q^2) = \frac{e^2}{16 \pi} k\,|\vec{p}_\pi| \frac{4 |\ampV^{({\rm u})}(q^2,s)|^2}{m_\pi^2}. \label{eq:xsection}
\end{align}
The cross section can be measured at $q^2=0$, i.e., with a real photon. In Eq.~(\ref{eq:xsection}), $\ampV^{({\rm u})}$ denotes the $\pi^+\gamma\to\pi^+\pi^0$ amplitude with the unsymmetrized $\pi^+\pi^0$ final state. Because
\begin{align}
  \vert \pi^+ (\pvec_1) \, \pi^0 (\pvec_2) \rangle &= \frac{1}{\sqrt{2}}\,\left( 
\vert \pi^+\pi^0 (\pvec_1, \pvec_2) ,\,I=1 \rangle \right. \cr
&\left. \hspace{6ex} + \vert \pi^+\pi^0 (\pvec_1, \pvec_2) ,\,I=2\rangle
\right), \nonumber
\end{align}
this amplitude is related to the amplitude with the isospin-projected final state through
\begin{align}
\nonumber  \ampV^{({\rm u})}  &= \frac{1}{\sqrt{2}} \ampV \,.
\end{align}
 
A second physically observable quantity is related to the $\rho$ resonance, which appears in the $\pi\pi$ system. The $\rho$ radiative decay width $\Gamma(\rho\to\pi\gamma)$ is determined by the photocoupling $G_{\rho\pi\gamma}=F(0, m_R^2 - \I m_R \Gamma_R)$ as \cite{Patrignani:2016xqp}
\begin{align}
&\Gamma(\rho \to \pi \gamma) =\frac{4}{3}\alpha \left(\frac{(m_\rho^2 - m_\pi^2)}{2m_\rho}\right)^3  \frac{|G_{\rho\pi\gamma}|^2}{m_\pi^2}. \label{eq:radwidth}
\end{align}
Note that the form factor $\formF(q^2,s)$, and hence the photocoupling $G_{\rho\pi\gamma}$, do not depend on whether one considers the unsymmetrized or the isospin-projected two-pion state. In the unsymmetrized case, Eqs.~(\ref{eq:T_BW_d}) and (\ref{eq:V_BW_d}) would read \cite{Briceno:2014uqa}
\begin{align}
&\ampT^{({\rm u})}(s) = \frac{8 \pi \sqrt{s}}{k} \frac{1}{\cot{\delta(s)} - \I} \nonumber
\end{align}
and
\begin{align}
\ampV^{({\rm u})}(q^2,s) & =  \sqrt{\frac{8 \pi}{k\Gamma(s)}} \frac{\formF(q^2,s) }{\cot{\delta(s) - \I}},  \nonumber
\end{align}
with the same $F(q^2,s)$.

%%%%%%%%%%%%%%%%%%%%%%%%%%%%%%%%%%%%%%%%%%%%%%%%%%%%%%%%%%%%%%%%%%%
\section{LATTICE PARAMETERS}\label{sec_lattice}
%%%%%%%%%%%%%%%%%%%%%%%%%%%%%%%%%%%%%%%%%%%%%%%%%%%%%%%%%%%%%%%%%%%

This calculation is performed on a single ensemble of gauge-field configurations with $2+1$ flavors of dynamical
clover fermions. This is the same ensemble as used in our calculation of $\pi\pi$ scattering \cite{Alexandrou:2017mpi},
and we refer the reader to that reference for further details. The main parameters are summarized in Table \ref{tab:lattice}.
The strange-quark mass is consistent with its physical value as determined via the ``$\eta_s$'' mass \cite{Davies:2009tsa, Dowdall:2011wh}.
The lattice scale was determined from the $\Upsilon$ $(1S)$-$(2S)$ splitting \cite{Davies:2009tsa,Meinel:2010pv},
where NRQCD \cite{Lepage:1992tx} with the physical $b$-quark mass was used to calculate the masses.
The renormalization factor $Z_V$ of the local vector current was determined by the LHPC Collaboration as explained in Ref.~\cite{Green:2017keo}.

\begin{table}[htb!]
\begin{tabular}{|r|c|}
\hline
    &   \texttt{C13}    \cr
\hline
  $N_L^3 \times N_T$  &   $32^3 \times 96 $     \cr
  $\beta$             &   $6.1$    \cr
  $N_f$               &   $2+1$                 \cr
  $c_{sw}$            &   $1.2493097$    \cr
  $a m_{u,d}$         &   $-0.285$    \cr
  $a m_{s}$           &   $-0.245$    \cr
  $N_{config}$        &   $1041$     \cr
  $a$ [fm]            &   $0.11403(77)$         \cr
  $L$ [fm]            &   $3.649(25)$                 \cr
  $am_{\pi}$          &   $0.18295(36)$          \cr
  $am_{N}$            &   $0.6165(23)$          \cr
  $am_{\eta_s}$       &   $0.3882(19)$          \cr
  $m_{\pi}L$          &   $5.865(32)$           \cr
  $Z_V$            &   $0.7903(2)$         \cr
\hline
\end{tabular}
\caption{The main parameters of the lattice gauge-field ensemble used in this work.  The uncertainties given here are statistical only.}
\label{tab:lattice}
\end{table}%%

%%%%%%%%%%%%%%%%%%%%%%%%%%%%%%%%%%%%%%%%%%%%%%%%%%%%%%%%%%%%%%%%%%%
\section{INTERPOLATING FIELDS AND CORRELATION FUNCTIONS}\label{sec_interpolators_and_3pt}
%%%%%%%%%%%%%%%%%%%%%%%%%%%%%%%%%%%%%%%%%%%%%%%%%%%%%%%%%%%%%%%%%%%

To determine the finite-volume matrix elements we are interested in, we need to compute
two-point functions for the single-pion system ($J^{PC}=0^{-+}$, $I=1$, $I_3=1$) and for the two-pion system  ($J^{PC}=1^{--}$, $I=1$, $I_3=1$), as well as
three-point functions with an insertion of the electromagnetic current. The generalized eigenvectors
obtained in the spectroscopic analysis of the two-point functions are then used to construct optimized three-point functions.

%-----------------------------------------------------------------%
\subsection{Two-point functions overlapping with \texorpdfstring{$\bm{ J^{PC}=0^{-+}}$}{JPC=0-+} }\label{subsec_pi}
%-----------------------------------------------------------------%

The projection of the single-pion field to an irreducible representation is trivial, i.e., it resides in the (pseudoscalar) $A_1^{(-)}$ irreducible representation \cite{Rummukainen:1995vs} for all momenta. For clarity we suppress the group indices of the single-pion field. We use the following interpolating operator:
\begin{align}
O_\pi^{\vecp} (\tSnk) = \sum_{\xvec} \, \bar{d}(\tSnk,\xvec)\,\gamma_5 \,u(\tSnk, \xvec) \,\epow{\I \vecp \cdot \xvec}\,,
\end{align}
with momentum $\vecp$. The associated correlator $C_\pi^{\vecp}$ is
\begin{align}
C_\pi^{\vecp}(t) = \langle  O_\pi^{\vecp}(\tSnk) O_\pi^{\vecp\dagger}(\tSnk - t) \rangle.
\end{align}
The ground-state contribution to the pion correlator, which is obtained in the limit of large $t$, has the decomposition
\begin{align}
C_\pi^{\vecp}(t) = \frac{Z_\pi^{\vecp} Z_\pi^{\vecp *}}{2E_\pi^{\vecp}} e^{-E_\pi^{\vecp} t},
\end{align}
where the overlap factor is defined as
\begin{align}
 \langle 0 | O_\pi^{\vecp} | \pi, \vecp \rangle_{FV} = Z_\pi^{\vecp}
\end{align}
and the finite-volume states are normalized such that
\begin{align}
 \langle \pi,\vecp^{\:\prime} | \pi, \vecp \rangle_{FV} = 2 E_\pi^{\vecp} \delta_{\vecp,\,\vecp^{\:\prime}}.
\end{align}
Because the pion is a stable hadron, its energy is affected only by exponentially suppressed finite-volume effects, which are negligible for our value of $m_\pi L$.
The dispersion relation of the pion was presented in Ref.~\cite{Alexandrou:2017mpi} and follows the relativistic form well.

%-----------------------------------------------------------------%
\subsection{Two-point functions overlapping with \texorpdfstring{$\bm{ J^{PC}=1^{--}}$}{JPC=0-+} }\label{subsec_pipi}
%-----------------------------------------------------------------%

The $J^{PC}=1^{--}$ two-point functions with momentum $\vecP$ are constructed using two types of interpolators, the single-hadron and the multi-hadron interpolators:
\begin{align}
  \label{eq:Oqq}
  O_{\qbar q} \big(t , \vec{P}\big) &=
  \sum_{\xvec} \, \bar{d}(t,\xvec)\,\Gamma_i \,u(t, \xvec) \,\epow{\I\vec{P}\cdot\xvec}\,,\\
%%%
    \nonumber  \label{eq:Opipi}
  O_{\pi\pi}\big(t, \pvec_1, \pvec_2\big) &=
    \frac{1}{\sqrt{2}} \,\big(\pi^+(t,\pvec_1)\, \pi^0(t,\pvec_2)\\
     &\qquad- \pi^0(t,\pvec_1)\,\pi^+(t,\pvec_2) \big)\, ,
\end{align}
where $\vec{p}_1+\vec{p}_2=\vec{P}$. To project these interpolators to definite irreps $\Lambda$ of the little group $LG(\vec{P})$,
we use the projection formulas with representation matrices \cite{Morningstar:2013bda}

\begin{equation}
\label{eq:qqprojector}
 O_{\qbar q}^{\vecP,\,\Lambda,\,r}(t) = \frac{\mathrm{dim}(\Lambda)}{N_{LG(\vec{P})}}
\sum_{ \hat{R} \in LG(\vec{P}) } \Gamma_{rr}^{\Lambda}(\hat{R}) \hat{R}\,O_{\qbar q}(t,\vec{P}),
\end{equation}
and
\begin{align}
\label{eq:projection_operator}
&O_{\pi\pi}^{\vecP,\,\Lambda,\,r}(t) = \frac{\mathrm{dim}(\Lambda)}{N_{LG(\vec{P})}} \cr
&\times\sum_{ \hat{R} \in LG(\vec{P}) }\Gamma_{rr}^{\Lambda}(\hat{R})\: O_{\pi\pi}\big(t,\, \vec{P}/2+\hat{R}\vec{p},\, \vec{P}/2-\hat{R}\vec{p}\,), \cr
\end{align}
where $\vec{p}$ takes on the values
\begin{align}
 \vec{p}=\frac{\vec{P}}{2} + \frac{2\pi}{L}\vec{m}, \;\;\vec{m}\in\mathbb{Z}^3.
\end{align}
Above, $\mathrm{dim}(\Lambda)$ is the dimension of the irrep, $N_{LG(\vec{P})}$ is the order of the Little Group $LG(\vecP)$,
and $\Gamma_{rr}^{\Lambda}(R)$ are suitably chosen representation matrices of $\hat{R} \in LG(\vec{P})$.
In our choice of basis indexing and projecting to finite-volume irreps, we use the $x,y,z$ polarization indices and
not the helicity basis like in Refs.~\cite{Moore:2005dw,Dudek:2012gj}.

In the following, we jointly denote the projected interpolators as
\begin{align}
O_i^{\vecP,\,\Lambda,\,r}(t),
\end{align}
where the index $i$ labels the type: $i=1$ and $i=2$
correspond to the quark-antiquark interpolators with $\Gamma_i=\gamma_i$ and $\Gamma_i=\gamma_0 \gamma_i$, respectively, and $i=3,4$ correspond to two-pion interpolators with different values of $|\vec{p}_1|$ and $|\vec{p}_2|$.

From the interpolators $O_i^{\vecP,\,\Lambda,\,r}$ we calculate a correlation matrix $C_{ij}^{\vecP,\,\Lambda,\,r}(t) = \langle  O_i^{\vecP,\,\Lambda,\,r}(\tSrc+t) O_j^{\vecP,\,\Lambda,\,r \dagger}(\tSrc) \rangle$. Its construction in terms of forward, sequential, and stochastic quark propagators is discussed in Ref.~\cite{Alexandrou:2017mpi}. The spectral decomposition of the two-point correlation matrix reads
\begin{align}
C_{ij}^{\vecP,\,\Lambda,\,r}(t) = \sum_n \frac{Z_i^{n,\,\vecP,\,\Lambda} Z_j^{n,\,\vecP,\,\Lambda\,\dagger}}{2E_n^{\vec{P},\Lambda}} e^{-E_n^{\vec{P},\Lambda} t}.
\end{align}
As in Sec.~\ref{subsec_pi}, we define the overlap factors as
\begin{align}
 \langle 0 | O_i^{\vecP,\,\Lambda,\,r} | n,\,\vecP,\,\Lambda,\,r \rangle_{FV} = Z_i^{n,\,\vecP,\,\Lambda},
\end{align}
and the finite-volume states are normalized such that
\begin{align}
 \langle n^\prime,\vecP^{\,\prime},\Lambda^\prime,r^\prime | n,\vecP,\Lambda,r \rangle_{FV} = 2 E_n^{\vec{P},\Lambda}  \delta_{n, n^\prime} \delta_{\vecP,\vecP^{\prime}} \delta_{\Lambda,\Lambda^\prime} \delta_{r,r^\prime}.
\end{align}
To extract the energies and overlap factors from the correlation matrix, we use the variational analysis \cite{Michael:1985ne, Luscher:1990ck, Orginos:2015tha, Blossier:2009kd} by solving the generalized eigenvalue problem
\begin{align}
&C_{ij}^{\vecP,\,\Lambda,\,r}(t) v_j^{n,\,\vecP,\,\Lambda}(t_0) = \lambda^{\vecP,\,\Lambda}_n(t,t_0) C_{ij}^{\vecP,\,\Lambda,\,r}(t_0) v_j^{n,\,\vecP,\,\Lambda}(t_0),
\end{align}
where we fix the normalization to
\begin{align}
\label{eq:v_norm}
&v_i^{n,\,\vecP,\Lambda \dagger}(t_0)\,C_{ij}^{\vecP,\,\Lambda,\,r}(t_0)\,v_j^{m,\,\vecP,\,\Lambda}(t_0) = \delta_{nm}.
\end{align}
Throughout this paper, we use the summation convention for
repeated indices $i$ or $j$. The principal correlators asymptotically behave as
\begin{align}
 \lambda^{\vecP,\,\Lambda}_n(t,t_0) = e^{-E_n^{\vec{P},\Lambda}(t - t_0)},
\end{align}
and we use single-exponential fits to extract $E_n^{\vec{P},\Lambda}$ \cite{Alexandrou:2017mpi}.

The generalized eigenvector $v_i^{n,\,\vecP,\,\Lambda}(t_0)$ can also be used to construct the optimized interpolator \cite{Michael:1985ne,Luscher:1990ck,Blossier:2009kd,Orginos:2015tha}
\begin{align}
\label{eq:opt_interpolator}
{\cal O}^{n,\,\vecP,\,\Lambda,\, r}(t, t_0) = v_i^{n,\,\vecP,\,\Lambda \, \dagger}(t_0)\, O_i^{\vecP,\,\Lambda,\, r}(t),
\end{align}
which has a dominant overlap to a single well-defined state labeled with $(n,\,\vec{P},\,\Lambda,\,r)$. Note that, although we perform the analysis independently for different rows $r$ of the irrep $\Lambda$, in the infinite-statistics limit the energies and eigenvectors are independent of $r$.

An important quantity related to the energy $E_n^{\vec{P},\Lambda}$ of the $\pi\pi$ system in the moving frame $\vecP$ is the invariant mass
\begin{align}
\label{eq:invmass}
\sqrt{s_n^{ \vec{P},\Lambda}}  = \sqrt{(E_n^{ \vec{P},\Lambda})^2 - \vec{P}^2},
\end{align}
which is also used to define the scattering momentum $k_n^{ \vec{P},\Lambda}$ via
\begin{align}
\label{eq:scattmom}
\sqrt{s_n^{ \vec{P},\Lambda}} = 2 \sqrt{m_{\pi}^2 + (k_n^{ \vec{P},\Lambda})^2}.
\end{align}
%

%-----------------------------------------------------------------%
\subsection{The three-point functions}\label{subsec_wick}
%-----------------------------------------------------------------%

The current insertion that represents the interactions between the photon and the hadrons depends on the photon momentum $\vec{q}$, which combined with the initial and final state momenta satisfies momentum conservation: $\vec{P}+\vec{q}-\vec{p}_\pi = 0$.
For the current insertion operator we use
\begin{align}
J_\mu(\tCurr,\vec{q}) = \sum_{\vec{x}} e^{\I \vec{q}\cdot{\vec{x}}} J_\mu(\tCurr,\vec{x}),
\end{align}
with the local current
\begin{align}
J_\mu(\tCurr,\vec{x}) = Z_V \,\Big(&\frac{2}{3} \bar{u}(\tCurr,\vec{x})\gamma_\mu u(\tCurr,\vec{x}) \cr
- &\frac{1}{3} \bar{d}(\tCurr,\vec{x})\gamma_\mu d(\tCurr,\vec{x}) \Big).
\end{align}
The renormalization coefficient $Z_V$ was determined in Ref.~\cite{Green:2017keo} and is listed in Table \ref{tab:lattice}.

\begin{figure*}[!htb]
  \centering
  \includegraphics[width=0.3\textwidth]{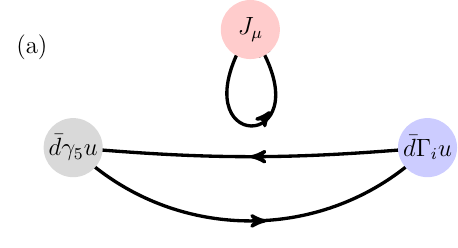}%
  \includegraphics[width=0.3\textwidth]{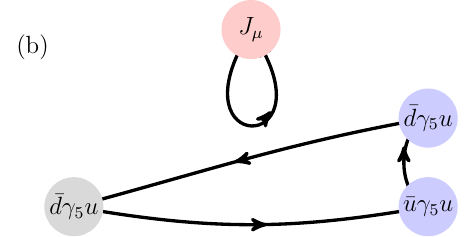}%  
  \includegraphics[width=0.3\textwidth]{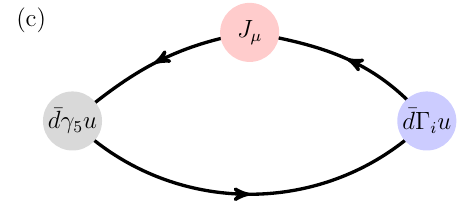}
  \includegraphics[width=0.3\textwidth]{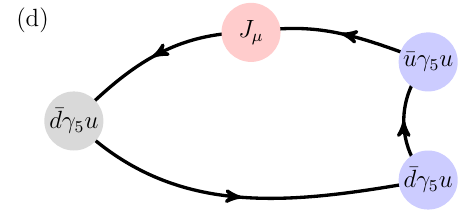}%
  \includegraphics[width=0.3\textwidth]{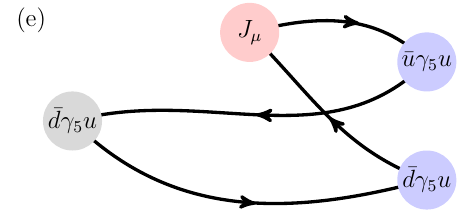}%
  \includegraphics[width=0.3\textwidth]{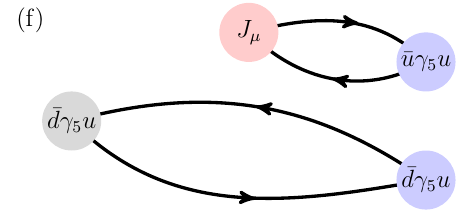}
  \caption{\label{fig:wick3pt} The different topologies of Wick contractions that make up the three-point function $C_{3,\;\mu,i}^{\vecp,\,\vecP,\,\Lambda,\,r}(\tSnk,\tCurr,\tSrc)$.}
\end{figure*}

The three-point correlation functions are then obtained from the sink/source interpolators and current insertion as
\begin{align}
&C_{3,\;\mu,i}^{\vecp,\,\vecP,\,\Lambda,\,r}(\tSnk,\tCurr,\tSrc) = \cr
&\langle O_\pi^{\vecp}(\tSnk)\,J_\mu(\tCurr,\vecQ)\, O_i^{\vecP,\,\Lambda,\,r\,\dag}(\tSrc) \rangle,
\end{align}
where $\tSrc$ is the source time, $\tCurr$ is the current insertion time and $\tSnk$ is the sink time. The three-point function is expressed in terms of quark propagators by evaluating Wick contractions. Figure \ref{fig:wick3pt} shows the quark-flow diagrams needed to calculate the $C_{3,\;\mu,i}^{\vecp,\,\vecP,\,\Lambda,\,r}$ three-point functions. The current-disconnected diagrams labeled (a) and (b), i.e. the diagrams where the quark flow goes from the current $J_\mu$ directly back to the current $J_\mu$, are omitted in this study. For the nucleon electromagnetic form factors, the current-disconnected contributions are known to be of order $1\%$ for the quark masses used here \cite{Green:2015wqa}.

The Wick contractions depicted in Fig.~\ref{fig:wick3pt} are constructed from point-to-all, sequential and stochastic time-slice propagators. The technique builds upon and extends the scheme used in Ref.~\cite{Alexandrou:2017mpi} for the construction of two-point correlation functions. This combination of propagator types allows for a compromise in flexibility to construct all required diagrams, minimal input of stochastic noise into correlation functions and economy in the cost of producing quark propagators and contractions.\\

The point-to-all propagators are obtained from the inversion of the Dirac operator $D$ on a fully spin- and color-diluted point-source localized at fixed source location $y$,
\begin{align}
  S(x; y)^{ab}_{\alpha\beta} &= \sum_z D^{-1}(x,z)^{ac}_{\alpha\gamma}\,\eta^{(y,b,\beta)}(z)^{c}_{\gamma}\, ,
  \label{eq:point_to_all_prop} \\
  %%%
  \eta^{(y,b,\beta)}(x)^b_\beta &= \delta_{x,y}\,\delta^{ab}\,\delta_{\alpha\beta} \,,\quad a,b = 0,1,2, \,\alpha,\beta = 0,1,2,3\,.
  \nonumber
\end{align}

The sequential propagator results from performing an additional inversion of the Dirac operator on a point-to-all propagator on sequential source time slice $t$ with insertion of a spin matrix $\Gamma$ and momentum $\pvec$ according to
\begin{align}
  T(x; t, \Gamma,\pvec; y)^{ab}_{\alpha\beta} &= \sum\limits_{\zvec}\,S(x; t, \zvec)^{ac}_{\alpha\gamma}\,\Gamma_{\gamma\delta}\,\epow{i\pvec \cdot \zvec}\,S(t,\zvec; y)^{cb}_{\delta\beta}\,.
  \label{eq:sequenial_prop}
\end{align}
For the purpose of this work the sequential source time slice always coincides with the source time slice , $t = t_y=y_0$ and $\Gamma = \gamma_5$ for the pseudoscalar vertex.\\

Finally, the stochastic propagators follow from inverting the Dirac matrix on stochastic time slice sources, whose components 
on a fixed time slice $t_y$ are independently and identically set with $\mathbb{Z}_2 + i \,\mathbb{Z}_2$ noise,
\begin{align}
  S(x,y)^{ab}_{\alpha\beta} &= \mathrm{E}\left[ \phi^{(t_y)}(x)^a_{\alpha}\,\xi^{(t_y)}(y)^{b\,*}_{\beta}  \right]\, ,
  \label{eq:stochastic_prop}\\
  %%%
  \xi^{(t_y)}(x)^b_{\beta} &= \delta_{t_x, t_y}\,\xi^{(t_y)}(\xvec)^b_{\beta} \,
  \in\, \left\{ \pm \frac{1}{\sqrt{2}} \pm \frac{i}{\sqrt{2}} \right\} \,\forall \, \xvec,\beta,b\, ,
  \label{eq:stochastic_source}\\
  %%%
  \phi^{(t_y)}(x)^a_{\alpha} &= \sum_z D^{-1}(x,z)^{ab}_{\alpha\beta}\,\xi^{(t_y)}(z)^b_{\beta}\,,
\end{align}
such that we have the expectation values
\begin{align}
  \mathrm{E}\,\left[ \xi^{(t)}(x)^a_{\alpha} \right] &= 0\, ,
  \label{eq:expval_xi}\\
  %%%
  \mathrm{E}\,\left[ \xi^{(t)}(x)^a_{\alpha} \, \xi^{(t)}(y)^{b\,*}_{\beta} \right]
  &= \delta_{t, t_x}\,\delta_{t_x, t_y}\,\delta_{\xvec, \yvec}\,\delta^{ab}\,\delta_{\alpha\beta}\,.
  \label{eq:expval_xixi}
\end{align}
As a variant of the stochastic time-slice propagator defined in Eq. (\ref{eq:stochastic_prop}) the one-end-trick based on spin diluted stochastic time-slice sources
is used to construct diagrams (e) and (f) in Fig.~\ref{fig:wick3pt}. The sources in Eq. (\ref{eq:stochastic_source}) are thus modified according to
\begin{align}
  \xi^{(t_y,\pvec,\lambda)}(x)^b_{\beta} &= \delta_{t_y, t_x}\,\delta_{\lambda\beta}\,\xi^{(t_y)}(\xvec)^b\,\epow{i\pvec \cdot \xvec}\,,\quad \lambda = 0,1,2,3\,.
  \label{eq:stochastic_eot_source}
\end{align}
The one-end-trick then allows for the representation of a product of quark propagatos by two stochastic propagators
through a vertex given again by $\Gamma$ and $\pvec$ as
\begin{align}
  & \mathrm{E}\left[ 
     \phi^{(t_y,\pvec, \kappa)}(x)^a_{\alpha} \,\left( \Gamma\, \gamma_5\right)_{\kappa\lambda}\,\phi^{(t_y,0,\lambda)}(z)^{b *}_{\beta'}\,\left( \gamma_5 \right)_{\beta'\beta}
  \right] 
  \label{eq:oet_propagator_product} \\
%%%   %%%
  & \quad = S(x; t_y, \yvec)^{ac}_{\alpha\kappa}\,\epow{i\pvec \cdot \yvec}\,\left( \Gamma\, \gamma_5\right)_{\kappa\lambda}\,S(z; t_y, \yvec)^{bc\, *}_{\beta'\lambda}\,\left( \gamma_5 \right)_{\beta'\beta} 
  \nonumber\\
  %%%
  & \quad = \left( S(x; t_y, \yvec)\,\epow{i\pvec\cdot\yvec}\, \Gamma\, S(t_y, \yvec; z) \right)^{ab}_{\alpha\beta}\,.
  \nonumber
\end{align}
Equation (\ref{eq:oet_propagator_product}) used in addition $\gamma_5$-hermiticity for the Dirac propagator, $S(x;y)^\dagger = \gamma_5\,S(y;x)\,\gamma_5$.

The quark propagator loops of the connected diagrams (c) and (d) are closed using the stochastic time-slice propagator from current vertex $J_\mu$ to pion vertex $\dbar\,\gamma_5\,u$
at sink. Based on the application of point-to-all and stochastic propagator these diagrams are factorized into elementary contractions. For diagram (c),
we have
\begin{align}
  & \Tr{ S(x_i; x_J )\,\gamma_\mu\,S(x_J; x_f)\,\gamma_5\,S(x_f; x_i)\,\Gamma_i }
  \label{eq:triangle_factors}\\
  &\quad = \mathrm{E} \left[ \eta_{\phi}(x_J)^a_{\alpha} \, \eta_{\xi}(x_f)^b_\beta  \right] \,
  \delta^{ab}\,\left( \Gamma_i\,\gamma_5 \right)_{\beta \alpha} \, ,
  \nonumber\\
  %%%
  & \eta_{\phi}(x_J) = S(x_J; x_i)^\dagger\,\gamma_5\gamma_\mu\,\phi^{(\tSnk)}(x_J) \, ,
  \nonumber\\
  %%%
  & \eta_{\xi}(x_f) = \xi^{(\tSnk)}(x_f)^\dagger\,\gamma_5\,S(x_f; x_i),
  \nonumber
\end{align}
where $\eta_{\phi,\xi}$ are contracted, Fourier transformed and stored separately as $\eta_{\phi}(\tCurr, \qvec)$ and $\eta_{\xi}(\tSnk, \vecp)$ for each stochastic sample. Subsequently they are used to recombine the diagram for all required momenta $\qvec,\,\vecp$ as well as any vertex $\Gamma_i$ and $\vecP$ at the source.
Diagram (d) follows analogously by promoting the point-to-all propagator $S(x_f; x_i)$ in Eq. (\ref{eq:triangle_factors}) to a sequential propagator,
\begin{align}
  & \Tr{ S(x_{i_1}; x_J )\,\gamma_\mu\,S(x_J; x_f)\,\gamma_5\,S(x_f; x_{i_2})\,\times \cr
  &\gamma_5\,\,\epow{i\pvec_{i_2}\xvec_{i_2}}\,S(x_{i_2}; x_{i_1})\,  \Gamma_i }
  \label{eq:box_factors}\\
  &\quad = \mathrm{E} \left[ \eta_{\phi}(x_J)^a_{\alpha} \, \eta_{\xi}(x_f)^b_\beta  \right] \,
      \delta^{ab}\,\left( \Gamma_i\,\gamma_5 \right)_{\beta \alpha} \, ,
  \nonumber\\
  %%%
  & \eta_{\phi}(x_J) = S(x_J; x_{i_1})^\dagger\,\gamma_5\gamma_\mu\,\phi^{(\tSnk)}(x_J) \, ,
  \nonumber\\
  %%%
  & \eta_{\xi}(x_f) = \xi^{(\tSnk)}(x_f)^\dagger\,\gamma_5\,T(x_f; \tSrc,\gamma_5, \pvec_{i_2}; x_{i_1})\,.
  \nonumber
\end{align}

For diagram (e) in Fig.~\ref{fig:wick3pt}, the one-end-trick setup in Eq. (\ref{eq:oet_propagator_product}) leads to the factorization of the diagram,
\begin{align}
  & \Tr{ S(x_{i_1}; x_J )\,\gamma_\mu\,S(x_J; x_{i_2})\,\epow{i\pvec_{i_2}\xvec_{i_2}}\, \cr
  &\times\gamma_5\,S(x_{i_2}; x_{f})\,\gamma_5\,\,S(x_{f}; x_{i_1})\,  \Gamma_i }
  \label{eq:z_factors}\\
  & \quad =
   \mathrm{E} \left[ \eta_{\phi}^{(\lambda)}(x_J)_\alpha \,\eta_{\bar{\phi}}^{(\lambda)}(x_f)_{\beta} \right] \,
     \left( \Gamma_i\,\gamma_5 \right)_{\beta\alpha}\, ,
   \nonumber\\
   %%%
   & \eta_{\phi}^{(\lambda)}(x_J) = S(x_J; x_{i_1})^\dagger\,\gamma_5 \gamma_\mu\,\phi^{(\tSrc, \pvec_{i_2}, \lambda)}(x_J) \, ,
   \nonumber\\
   %%%
   & \eta_{\bar{\phi}}^{(\lambda)}(x_f) = \phi^{(\tSrc,0,\lambda)}(x_f)^\dagger\,S(x_f; x_{i_1})\,.
   \nonumber
\end{align}
Finally, diagram (f) is calculated as the product of propagator loop traces using again the one-end-trick,
\begin{align}
  & \Tr{ S(x_{i_2}; x_J )\,\gamma_\mu\,S(x_J; x_{i_2}) \,\gamma_5}\,\epow{i\pvec_{i_2}\xvec_{i_2}}
  \nonumber\\
  & \qquad \times \Tr{ S(x_{i_1}; x_{f})\,\gamma_5\,\,S(x_{f}; x_{i_1})\,  \Gamma_i }
  \label{eq:direct_factors}\\
  %%%
  & \quad = \mathrm{E}\left[ \phi^{(\tSrc,0,\lambda)}(x_J)^{\dagger}\,\gamma_5\gamma_\mu\,\phi^{(\tSrc,\pvec_{i_2}, \lambda)} \right]
    \nonumber\\
  & \qquad \times  \Tr{S(x_{f}; x_{i_1})^{\dagger}\,S(x_f; x_{i_1})\,\Gamma_i\,\gamma_5}
    \nonumber
\end{align}
All quark propagators are smeared at their source and sink side in the same way as in Ref.~\cite{Alexandrou:2017mpi}, except the end of propagators joining the local current insertion vertex.

%-----------------------------------------------------------------%
\subsection{Optimized three-point functions}\label{subsec_optimized}
%-----------------------------------------------------------------%

The spectral decomposition of the three-point function $C_{3,\;\mu,i}^{\vecp,\,\vecP,\,\Lambda,\,r}(\tSnk,\tCurr,\tSrc)$, keeping as before only the ground-state contribution for the pion (for large $\tSnk-\tCurr$), is
\begin{align}
\label{eq:C3_decomp}
& C_{3,\;\mu,i}^{\vecp,\,\vecP,\,\Lambda,\,r}(\tSnk,\tCurr,\tSrc) \cr
&  =\sum_{n} Z_\pi^{\vecp}\: Z_i^{n,\,\vecP,\,\Lambda\,\dag}\: \langle \pi,\vecp | J_\mu(0,\vecQ)|n,\vecP,\,\Lambda,\,r\rangle_{FV}\cr
&\qquad \times \frac{e^{-E_\pi^{\vecp} (\tSnk - \tCurr)} e^{-E_n^{\vecP, \Lambda} (\tCurr - \tSrc)} }{2E_n^{\vecP, \Lambda } 2E_\pi^{\vecp}}.
\end{align}
For the $\pi\pi$ system we want to project to the $n$-th state. This will allow us to have a definite invariant mass, $\sqrt{s_n^{\vecP,\Lambda}}$, and momentum transfer,
$(q^2)_{n,\vecP,\Lambda}^{\vecp} = (E_n^{\vecP,\Lambda} - E_\pi^{\vecp})^2 - \vecQ^2$, in our matrix element. To achieve this we utilize the orthogonality between the generalized eigenvectors and overlap factors\footnote{Note that this choice depends on the normalization of the generalized vectors [cf.~Eq.~\eqref{eq:v_norm}].},
\begin{align}
v_i^{n\,\vecP, \Lambda}(t_0)\, Z_i^{m\,\vecP, \Lambda\,\dagger} = \sqrt{2E_n^{\vecP, \Lambda}}\,e^{E_n^{\vecP, \Lambda} t_0 / 2} \delta_{nm},
\end{align}
and construct the optimized three-point function \cite{Dudek:2009kk,Becirevic:2014rda,Shultz:2015pfa}
\begin{align}
&\Omega_{3,\;\mu,\, n}^{\vecp,\,\vecP,\,\Lambda,\,r}(\tSnk,\tCurr,\tSrc, t_0) \cr
&=  v_i^{n\,\vecP, \Lambda}(t_0)\:C_{3,\;\mu,i}^{\vecp,\,\vecP,\,\Lambda,\,r}(\tSnk,\tCurr,\tSrc) \cr
&=\langle O_\pi^{\vecp}(\tSnk)\,J_\mu(\tCurr,\vecQ)\, {\cal O}^{n,\,\vecP,\,\Lambda,\, r}(\tSrc, t_0) \rangle.
\end{align}
This gives
\begin{align}
\label{eq:omega}
&\Omega_{3,\;\mu,\, n}^{\vecp,\,\vecP,\,\Lambda,\,r}(\tSnk,\tCurr,\tSrc, t_0) \cr
&= \sqrt{2E_n^{\vecP, \Lambda}}\, e^{E_n^{\vecP, \Lambda} t_0 / 2} \, Z_\pi^{\vecp} \, \langle \pi,\vecp |J_\mu(0,\vecQ) | n, \vecP, \Lambda,r \rangle_{FV}\cr
&\qquad\times\frac{ e^{-E_\pi^{\vecp} (\tSnk - \tCurr)} e^{-E_n^{\vecP, \Lambda} (\tCurr - \tSrc)}}{2E_n^{\vecP, \Lambda} 2E_{\pi}^{\vecp}},
\end{align}
and we see that the optimized three-point function overlaps only to the single definite state $|n; \vecP, \Lambda,r\rangle$.

%%%%%%%%%%%%%%%%%%%%%%%%%%%%%%%%%%%%%%%%%%%%%%%%%%%%%%%%%%%%%%%%%%%
\section{DETERMINING THE FINITE-VOLUME MATRIX ELEMENTS}\label{sec_matrix_elements}
%%%%%%%%%%%%%%%%%%%%%%%%%%%%%%%%%%%%%%%%%%%%%%%%%%%%%%%%%%%%%%%%%%%
% 
\begin{figure*}
\includegraphics[width=\textwidth]{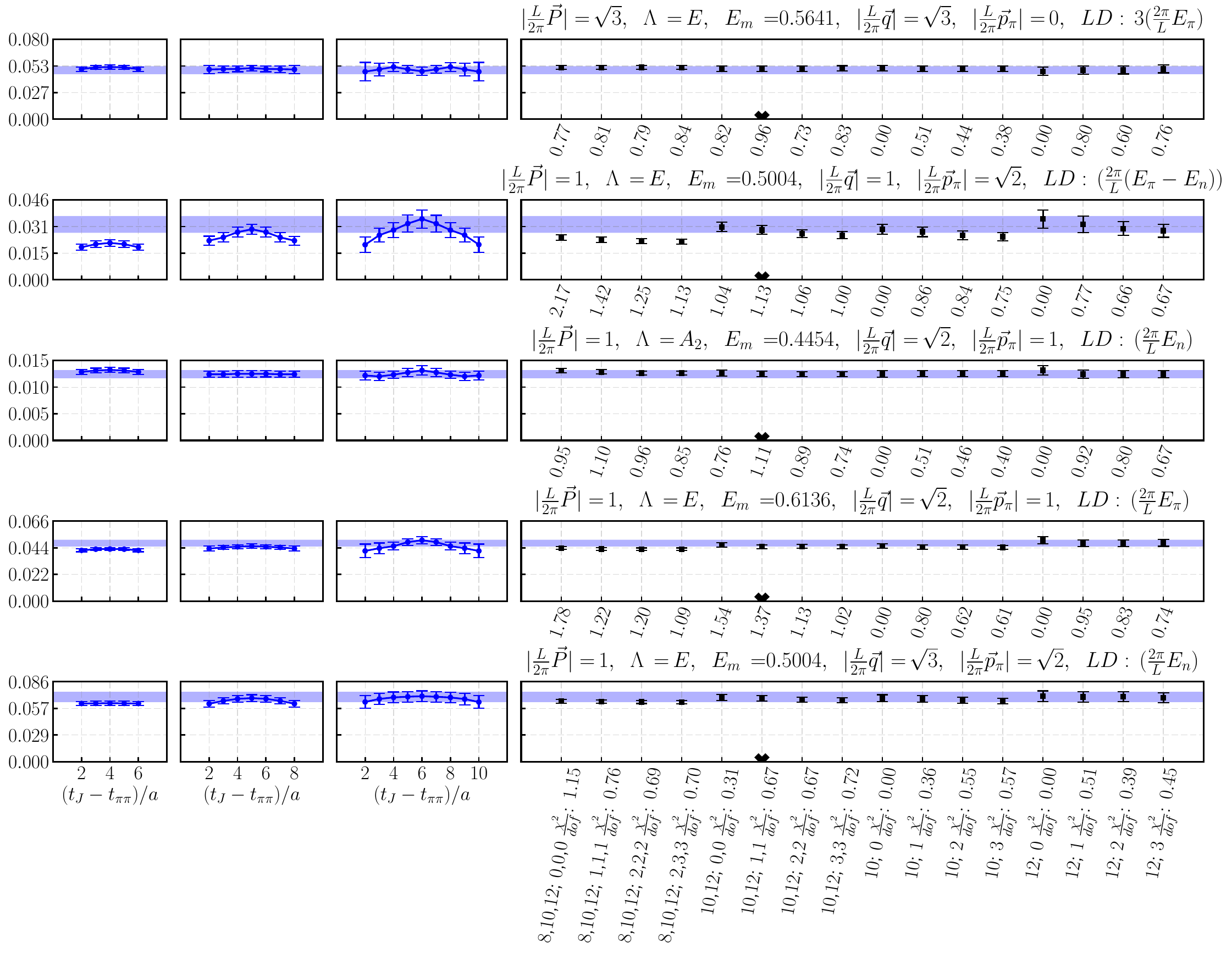}
\caption{\label{fig:example_ME_fit}Examples of results for the finite-volume matrix elements  $|\langle \pi,\vecp | J_\mu(0,\vecQ)|n,\vecP,\,\Lambda,\,r\rangle_{FV}|$.
The left three panels show the data as a function of $\tCurr-\tSrc$ for the three different source-sink separations. The right panels show the fitted values for multiple different fit ranges, which are indicated at the bottom. There, the first set of numbers are the included source-sink separations, and the second set of numbers are the distances
from the mid-point that are included for each of these source-sink separations. The blue bands show the chosen fit result, and the half-crosses mark the fits that are used to estimate systematic uncertainties. The values of $\chi^2/{\rm dof}$ are also given. The quantity denoted as $LD$ is the kinematic factor appearing next to $2i \ampV/m_\pi$ in Eq.~(\ref{eq:LLdecomp}).}
\end{figure*}
To extract the finite-volume matrix elements $\langle \pi,\vecp | J_\mu(0,\vecQ)|n,\vecP,\,\Lambda,\,r\rangle_{FV}$ from the correlation functions, we construct the ratio
\begin{align}
\label{eq:ratio}
& R_{\mu,\, n}^{\vecp,\,\vecP,\,\Lambda,\,r}(\tSnk,\tCurr,\tSrc) \cr
& = \frac{\Omega_{3,\;\mu,\, n}^{\vecp,\,\vecP,\,\Lambda,\,r}(\tSnk,\tCurr,\tSrc, t_0)\: \Omega_{3,\;\mu,\, n}^{\vecp,\,\vecP,\,\Lambda,\,r\,\dag}(\tSnk, t^\prime,\tSrc, t_0)}{C_{\pi}^{\vecp}(\DeltaT)\:\lambda_n^{\vecP, \Lambda}(\DeltaT,t_0)}, \cr
\end{align}
where $C_\pi^{\vecp}$ is the pion correlator, $\lambda_n^{\vecP, \Lambda}$ is the principal correlator of the variational analysis, $\DeltaT = \tSnk - \tSrc$ is the source-sink separation, and $t^\prime = \tSrc + \tSnk - \tCurr$.
The $t_0$ dependence of the optimized three-point function cancels with the $t_0$ dependence of the principal correlator. Inserting Eq.~\eqref{eq:omega} into Eq.~\eqref{eq:ratio} gives (for large time separations)
\begin{align}
\label{eq:ratio_ME}
R_{\mu,\, n}^{\vecp,\,\vecP,\,\Lambda,\,r}(\tSnk,\tCurr,\tSrc) = \frac{|\langle \pi,\vecp | J_\mu(0,\vecQ)|n,\vecP,\,\Lambda,\,r\rangle_{FV}|^2}{4 E_n^{\vecP, \Lambda} E_\pi^{\vecp}}.
\end{align}
The matrix elements determined from Eq.~\eqref{eq:ratio} still contain residual excited-state contamination that decays exponentially for large $\DeltaT$, $\tCurr-\tSrc$, and $\tSnk-\tCurr$. We have data for $\DeltaT/a=8,10,12$. There are several ways to proceed from this point on:
\begin{itemize}
	\item[1)] Set $\tCurr - \tSrc = \DeltaT/2$ and fit only the $\DeltaT$ dependence of the matrix element with an excited-state model, as for example in Ref.~\cite{Detmold:2012ge},
	\item[2)] Fit both the $\DeltaT$ and $\tCurr-\tSrc$ dependence with an excited-state model,
	\item[3)] Fit constants to the ratios (assuming that only the desired initial and final states contribute), varying the time ranges to assess residual contamination.
\end{itemize}
We found that that options 1) and 2) did not yield stable fits, because we have too few source-sink separations and the statistical uncertainties are too large.
We therefore use option 3), where we investigate whether the various fits are statistically compatible, and estimate a systematic uncertainty associated with the fit choice.
In Fig.~\ref{fig:example_ME_fit} we present results for the matrix elements $|\langle \pi,\vecp | J_\mu(0,\vecQ)|n,\vecP,\,\Lambda,\,r\rangle_{FV}|$ at representative kinematic points (plots for the other kinematic points are shown in Appendix \ref{app:ME_fits}). As explained in the caption of the figure, we perform fits
for many different time ranges and then choose one that appears to have plateaued for the further analysis. To estimate the systematic uncertainty
associated with the fit range for the ratio, we compute the change in the central value when going from the chosen fit to $\DeltaT/a=10$, as marked with an $X$ in Fig.~\ref{fig:example_ME_fit}. As a cross-check, we also tested an alternative method for extracting the matrix elements, in which we did not use ratios, but fitted the three-point functions (\ref{eq:omega}) after dividing out the time dependence and overlap factors. That method gives results consistent with the ratio method. Because the ratio \eqref{eq:ratio_ME} also depends on the energies $E_{n}^{\vecP,\,\Lambda}$, we additionally include a second
systematic uncertainty associated with the choice of fit range used in the spectrum analysis of Ref.~\cite{Alexandrou:2017mpi}.
The numerical results for all kinematic points are listed in Tables \ref{tab:MEpt1} and \ref{tab:MEpt2} in Appendix \ref{app:ME_fits}. There, both systematic
uncertainties have been added in quadrature to the statistical uncertainties.

%%%%%%%%%%%%%%%%%%%%%%%%%%%%%%%%%%%%%%%%%%%%%%%%%%%%%%%%%%%%%%%%%%%
\section{MAPPING FROM FINITE VOLUME TO INFINITE VOLUME}\label{subsec_LLmap}
%%%%%%%%%%%%%%%%%%%%%%%%%%%%%%%%%%%%%%%%%%%%%%%%%%%%%%%%%%%%%%%%%%%

\subsection{Lellouch-L\"uscher factors}

\begin{figure*}
\includegraphics[width=\textwidth]{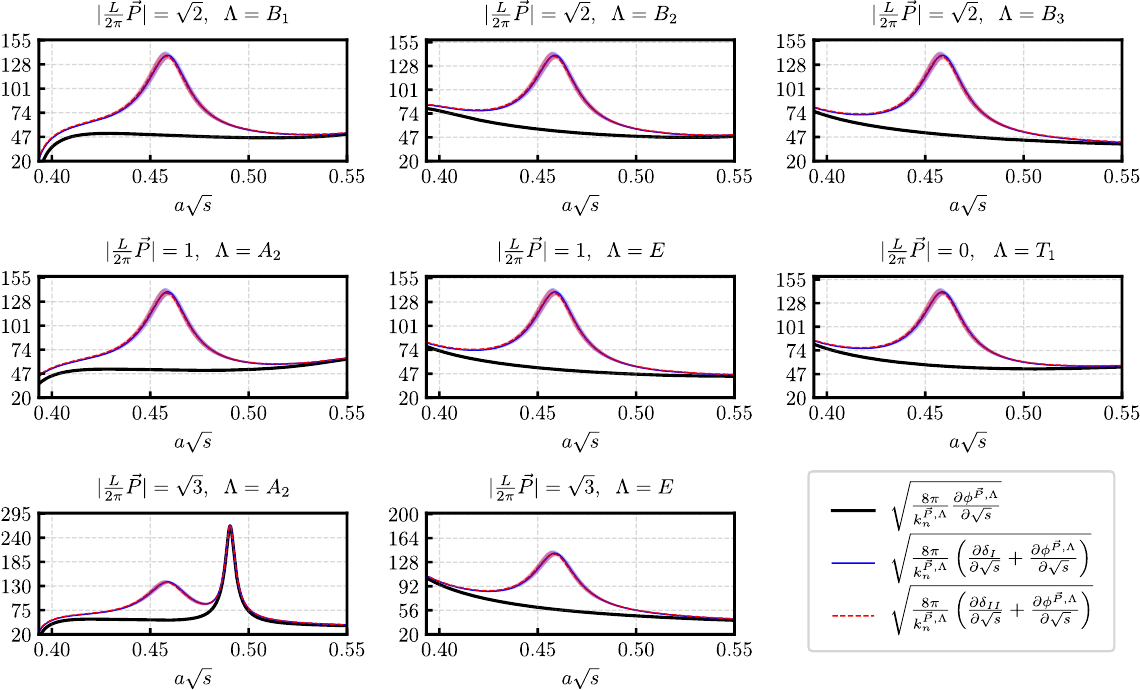}
\caption{\label{fig:LLmaps} The Lellouch-L\"uscher factors as a function of invariant mass, for the momentum frames and irreps used here. The thick black lines show the noninteracting Lellouch-L\"uscher factors (without the phase-shift derivative). The thin blue lines and dashed red lines show the full Lellouch-L\"uscher factors,
using the {\bf BW I} and {\bf BW II} models \cite{Alexandrou:2017mpi} for the scattering phase shift. The bands indicate the statistical uncertainties.}
\end{figure*}

The mapping between a finite-volume matrix element $|\langle \pi,\vecp | J_\mu(0,\vecQ)|n,\vecP,\,\Lambda,\,r\rangle_{FV}|$ calculated on the lattice  and the corresponding infinite-volume matrix element $|\langle \pi, \vec{p}_\pi |J_\mu(0) |  s, \vecP,\,\Lambda,\,r \rangle_{IV}|$, for our normalization of states, is \cite{Lellouch:2000pv, Briceno:2014uqa,Briceno:2015csa,Christ:2015pwa}
\begin{align}
\label{eq:LLmap}
& \frac{|\langle \pi, \vec{p}_\pi |J_\mu(0) |  s, \vecP,\,\Lambda,\,r \rangle_{IV}|^2}{|\langle \pi,\vecp | J_\mu(0,\vecQ)|n,\vecP,\,\Lambda,\,r\rangle_{FV}|^2} \cr & = \frac{1 }{2E_n^{\vecP,\Lambda}}    \frac{16 \pi \sqrt{s_n^{\vecP,\Lambda}}}{k^{\vecP,\Lambda}_n} \left( \frac{\partial \delta}{\partial E} + \frac{\partial \phi^{\vecP,\Lambda}}{\partial E} \right) \bigg|_{E=E_n^{\vecP,\Lambda}}. \qquad
\end{align}
Note that the current in the infinite-volume matrix element is evaluated in position space at $\vec{x}=0$, while the current in the finite-volume
matrix element is projected to momentum $\vec{q}$. The energy-dependence of the $\pi\pi$ $P$-wave scattering phase shift $\delta$ has to be determined from the L\"uscher analysis on the same lattice. We use our Breit-Wigner fits from Ref.~\cite{Alexandrou:2017mpi}, as already discussed in Sec.~\ref{sec_about_photproduction}. The function $\phi^{\vecP,\Lambda}$ in Eq.~(\ref{eq:LLmap}) appears in the L\"uscher quantization condition as
\begin{align}
\cot{\delta} = \cot{\phi^{\vecP,\Lambda}} = \sum_{l,m} c_{lm}^{\,\vecP,\Lambda} w_{lm}(k_{\vecP,\Lambda}^2),
\end{align}
where $w_{lm}$ is defined as
\begin{align}
w_{lm}(k^2) = \frac{ Z_{lm}^{\vecP} \left(1;(k L/(2 \pi))^2\right) }{ \pi^{3/2} \sqrt{2l+1} \gamma (\frac{k L}{2 \pi})^{l+1}},
\end{align}
with the generalized zeta function $Z_{lm}^{\vecP}$ and the Lorentz gamma factor $\gamma$. The quantization conditions for $\cot{\phi^{\vecP,\Lambda}}$ used are discussed in Sec.~VI of Ref~\cite{Alexandrou:2017mpi}; the nonzero factors $c_{lm}^{\vecP,\Lambda}$ appearing in elastic $P$-wave $\pi\pi$ scattering are also listed in Table \ref{tab:clmi}. 
\begin{table}[htb]
\begin{tabular}{|c|c|c|c|}
\hline
$\frac{L}{2\pi}\vecP$    & \hspace{1ex} $\Lambda$ \hspace{1ex}  & \hspace{1ex} $(l,m)$ \hspace{1ex} &  $c_{lm}^{\vecP,\Lambda}$    \cr
\hline
\hspace{1ex} $(0,0,0)$ \hspace{1ex} & $T_1$ & $(0,0)$ & $1$ \cr
\hline
$(0,0,1)$ & $A_2$ & $(0,0)$ & $1$ \cr
& & $(2,0)$ & $2$ \cr
\hline
 & $E$ & $(0,0)$ & $1$ \cr
& & $(2,0)$ & $-1$ \cr
\hline
$(0,1,1)$ & $B_1$ & $(0,0)$ & $1$ \cr
& & $(2,0)$ & $\frac{1}{2}$ \cr
& & $(2,1)$ & $\I\sqrt{6}$ \cr
& & $(2,2)$ & $-\sqrt{\frac{3}{2}}$ \cr
\hline
 & $B_2$ & $(0,0)$ & $1$ \cr
& & $(2,0)$ & $\frac{1}{2}$ \cr
& & $(2,1)$ & $-\I\sqrt{6}$ \cr
& & $(2,2)$ & $-\sqrt{\frac{3}{2}}$ \cr
\hline
 & $B_3$ & $(0,0)$ & $1$ \cr
& & $(2,0)$ & $-1$ \cr
& & $(2,2)$ & $\sqrt{6}$ \cr
\hline
$(1,1,1)$ & $A_2$ & $(0,0)$ & $1$ \cr
& & $(2,1)$ & $-\I\sqrt{\frac{8}{3}}$ \cr
& & $(2,2)$ & $-\sqrt{\frac{8}{3}}(\mathrm{Re}\, +\, \mathrm{Im})$ \cr
\hline
 & $E$ & $(0,0)$ & $1$ \cr
& & $(2,0)$ & $\I\sqrt{6}$ \cr
\hline
\end{tabular}
\caption{Nonzero values of $c_{lm}$ appearing in the quantization condition for elastic $P$-wave $\pi\pi$ scattering. Above, the term with $\mathrm{Re}$ and $\mathrm{Im}$ means $-\sqrt{\frac{8}{3}}(\mathrm{Re}[w_{22}]+\mathrm{Im}[w_{22}])$. }
\label{tab:clmi}
\end{table}%
The right-hand side of Eq.~\eqref{eq:LLmap}, known as the Lellouch-L\"uscher factor, depends on the $\pi\pi$ system's momentum $\vecP$, irreducible representation $\Lambda$, invariant mass $\sqrt{s_n^{\vecP,\Lambda}}$, and scattering momentum $k_n^{\vecP,\Lambda}$. In Fig.~\ref{fig:LLmaps} we show the Lellouch-L\"uscher factors as a function of invariant mass.

Calculating the derivative $\frac{\partial \phi^{\vecP,\Lambda}}{\partial E}$ in practice means that we must calculate the derivative of $w_{lm}(k^2)$:
\begin{align}
\frac{\partial \phi^{\vecP,\Lambda}}{\partial E} =\:& \frac{s^2 - (m_1^2-m_2^2)^2}{2\sqrt{s}^3}\frac{1}{1+\cot^2{\phi^{\vecP,\Lambda}}} \cr
&\times \sum_{l,m} c_{lm} \frac{\partial w_{lm}(k^2)}{\partial k^2},
\end{align}
where $m_1$, $m_2$ are the two hadron masses; in the case of $\pi\pi$ scattering $m_1=m_2=m_\pi$. In the rest frame, the
derivative of $Z_{lm}$ is again a zeta function:
\begin{align}
\frac{\partial}{\partial \hat{k}^2} Z_{lm}^{\vecP=\vec{0}}(s;\hat{k}^2) = s Z_{lm}^{\vecP=\vec{0}}(s+1;\hat{k}^2).
\end{align}
Since this does not hold in moving frames, we compute the derivative numerically.

In Fig.~\ref{fig:LLmaps} we can see that the two different models for the phase shift $\delta$, {\bf BW I} and {\bf BW II}, are statistically compatible.
Nevertheless, we use both Breit-Wigner models in our analysis to quantitatively assess this.

The fitting systematic uncertainties in $E_n^{\vecP,\Lambda}$ enter in the Lellouch-L\"uscher factors not only via the explicit factor of $E_n^{\vecP,\Lambda}$ in
Eq.~\eqref{eq:LLmap}, but also through the phase-shift parametrization fitted to these energies via the L\"uscher quantization condition. In Ref.~\cite{Alexandrou:2017mpi}, we estimated the systematic uncertainties in $E_n^{\vecP,\Lambda}$ by comparing the results of exponential fits with start times $t_{\rm min}$ and $t_{\rm min}+a$. To correctly propagate these uncertainties to the Breit-Wigner parameters, we then performed the L\"uscher analysis and the Breit-Wigner fits for both sets of energies \cite{Alexandrou:2017mpi}. In the present work, we therefore also repeat the mappings of the $\pi\gamma \to \pi\pi$ matrix elements (and the subsequent analysis) for both sets of Breit-Wigner parameters.

%-----------------------------------------------------------------%
\subsection{Lorentz decomposition of the infinite-volume matrix elements}\label{subsec_ME_IV}
%-----------------------------------------------------------------%

\begin{figure*}[htb]
\begin{center}
	\includegraphics[width=0.49\linewidth]{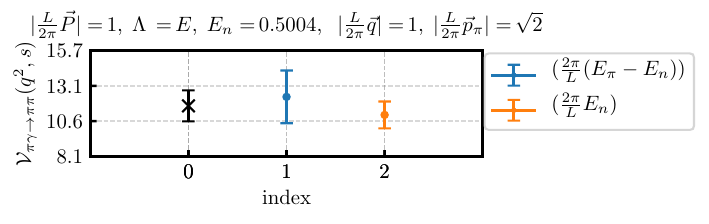}\hfill
	\includegraphics[width=0.49\linewidth]{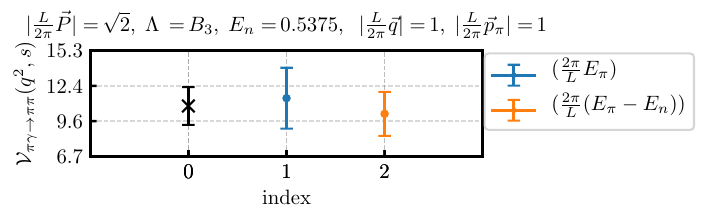}
	
	\vspace{4ex}
	
	\includegraphics[width=0.49\linewidth]{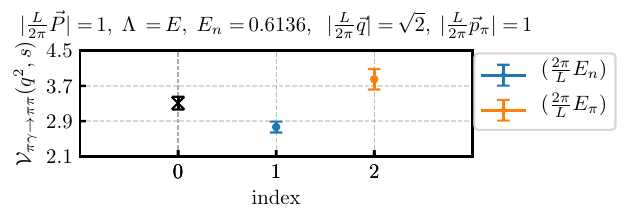}\hfill
	\includegraphics[width=0.49\linewidth]{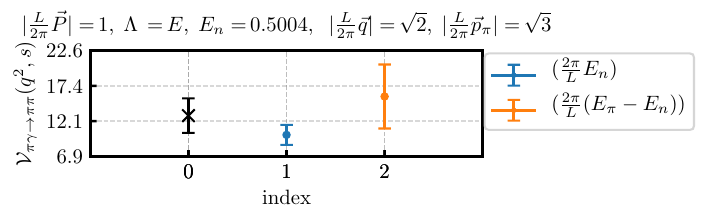}
	
	\vspace{4ex}
	
	\includegraphics[width=0.49\linewidth]{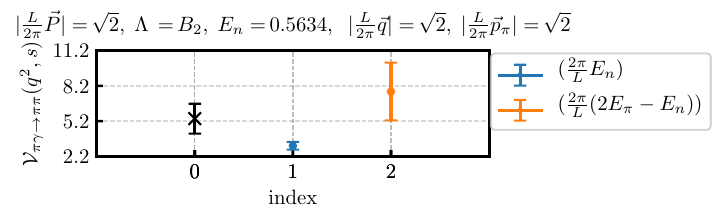}\hfill
	\includegraphics[width=0.49\linewidth]{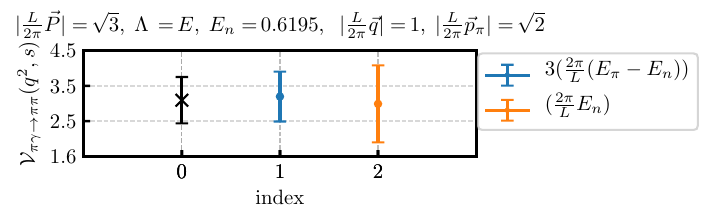}
	
	\vspace{4ex}
	
	\includegraphics[width=0.49\linewidth]{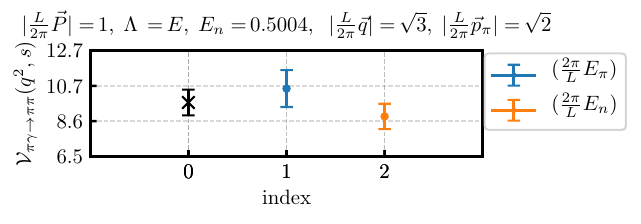}\hfill
	\includegraphics[width=0.49\linewidth]{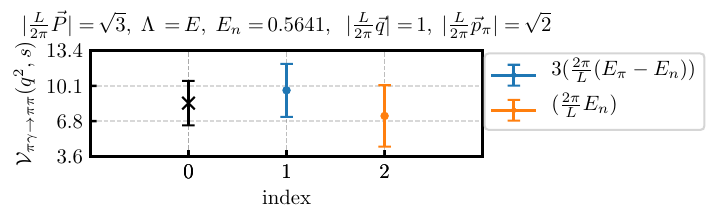}
	
	\vspace{4ex}

	\includegraphics[width=0.49\linewidth]{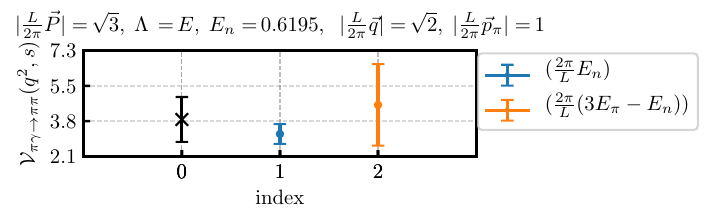}\hfill
	\includegraphics[width=0.49\linewidth]{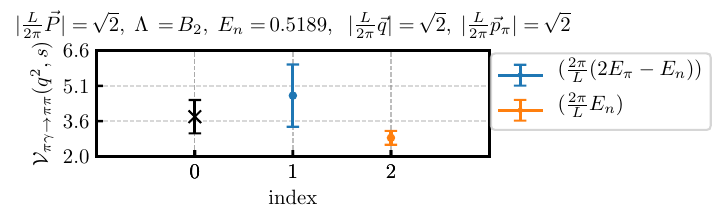}
	
	\vspace{4ex}

	\includegraphics[width=0.49\linewidth]{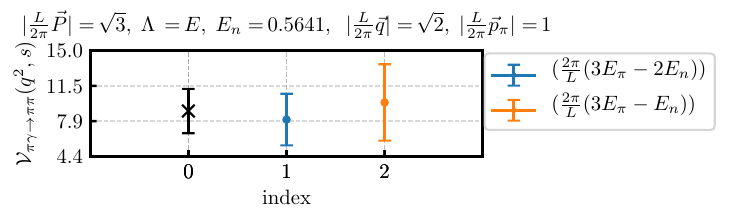}
	\caption{\label{fig:av_MEIV_II}  Transition amplitude values determined from the matrix elements $|\langle s, \vecP,\,\Lambda,\,r |J_\mu(0,\vecQ) | \pi, \vecp\rangle|_{IV}$, for those kinematic points $(s,q^2)$ where the choices of momentum directions and irrep indices yield two different Lorentz decomposition (LD) factors. For each kinematic point, we divide by the LD factors and average the two resulting values of $\ampV$. On the horizontal axes, the indices $1$ and $2$ correspond to the two different Lorentz decompositions indicated in the legend, while the index $0$ corresponds to the average. The data shown here are based on the {\bf BW II} Breit-Wigner parametrization; the results for {\bf BW I} look very similar.}
\end{center}
\end{figure*}

The infinite-volume matrix elements $\langle s, \vecP,\,\Lambda,\,r |J_\mu(0,\vecQ) | \pi, \vecp\rangle_{IV}$ obtained from Eq.~\eqref{eq:LLmap} still
carry the finite-volume irrep indices $\vecP$, $\Lambda$, $r$. The infinite-volume states $\langle s, \vecP,\,\Lambda,\,r |$ are linear combinations of the states labeled by the continuum polarization index $m$ in Eq.~\eqref{eq:IVstates}. The coefficients of these linear combinations are given by the irrep projection formula Eq.~\eqref{eq:qqprojector}. We form the same linear combinations of the polarization vectors on the right-hand side of Eq.~\eqref{eq:LLdecomp} to obtain the irrep-projected form-factor decompositions. Taking this into account, we can determine the values of the infinite-volume transition amplitude $\ampV$. Most kinematic points only have a single possible Lorentz-decomposition factor, but at certain values of $(s,q^2)$ there are two,
as shown in Fig.~\ref{fig:av_MEIV_II}. We average over the two resulting values of $\ampV$, which reduces the full set of $59$ matrix elements to $48$ distinct kinematic points $(s,q^2)$.

%%%%%%%%%%%%%%%%%%%%%%%%%%%%%%%%%%%%%%%%%%%%%%%%%%%%%%%%%%%%%%%%%%%
\section{Fitting the amplitude \texorpdfstring{$\bm{\ampV}$}{V}}\label{sec_multihadron}
%%%%%%%%%%%%%%%%%%%%%%%%%%%%%%%%%%%%%%%%%%%%%%%%%%%%%%%%%%%%%%%%%%%
%-----------------------------------------------------------------%
\subsection{Parametrization of the infinite-volume transition amplitude}\label{subsec_model}
%-----------------------------------------------------------------%

To allow the calculation of observables, the transition amplitude $\ampV(q^2,s)$ determined with lattice QCD at $48$ discrete values of $q^2$ and $s$ needs to be fitted to an analytic parametrization. In Sec.~\ref{sec_about_photproduction}, we factored out the $\rho$ pole in $s$ according to Watson's theorem,
\begin{align}
\ampV(q^2,s) = \frac{\formF(q^2,s)}{m_R^2 - s - \I\sqrt{s}\,\Gamma_i(s)}\sqrt{\frac{16\pi s \Gamma_i(s)}{k}}.
\end{align}
What remains is the transition form factor $\formF(q^2,s)$, which should not have any additional poles in $s$ in our region of interest.
To obtain a model-independent parametrization of $\formF(q^2,s)$, we perform a two-dimensional Taylor expansion in the variables
\begin{align}
{\cal S} = \frac{s - m_R^2}{m_R^2}  \label{eq:calS}
\end{align}
and
\begin{align}
z = \frac{\sqrt{t_+ - q^2} - \sqrt{t_+ - t_0}}{\sqrt{t_+ - q^2} + \sqrt{t_+ - t_0}}, \label{eq:z}
\end{align}
after dividing out the lowest expected pole in $q^2$:
\begin{align}
\formF(q^2,s) = \frac{1}{1 - \frac{q^2}{m_P^2}} \sum_{n,m} A_{nm} z^n \mathcal{S}^m. \label{eq:Fseries}
\end{align}
The variable ${\cal S}$ was chosen to be dimensionless and small near the resonance. The
definition of $z$ maps the complex $q^2$ plane, cut along the real axis for $q^2>t_+$, to the interior of the unit circle \cite{Boyd:1994tt,Boyd:1997qw,Bourrely:2008za,Bhattacharya:2011ah,Bhattacharya:2015mpa,Meyer:2016oeg}.
The constant $t_0$ determines which value of $q^2$ is mapped to $z=0$; we choose $t_0=0$. The constant $t_+$ should be set to the lowest branch point.
For the QED current, the branch cut starts at $(3 m_\pi)^2$ and the lowest pole is located at $m_\omega^2$. However, because we neglect the disconnected
contributions, we use $t_+=(2m_\pi)^2$ and $m_P=m_\rho$\footnote{Because $m_P^2 > t_+$, it is not actually necessary to factor out the pole, but there is no harm in doing so.}

In practice, the series (\ref{eq:Fseries}) needs to be truncated. We organize these truncations into three different families:
\begin{itemize}
 \item[{\bf F1)}] Combined order $K$: 
 \begin{align}
 \formF(q^2,s) = \frac{1}{1 - \frac{q^2}{m_P^2}} \sum_{n+m\leq K} A_{nm} z^n \mathcal{S}^m,
 \end{align}
 
 \item[{\bf F2)}] Order $N$ in $z$, combined order $K$:  
 \begin{align}
    \formF(q^2,s) = \frac{1}{1 - \frac{q^2}{m_P^2}} \sum_{n=0}^{N} \sum_{m=0}^{K-n} A_{nm} z^n \mathcal{S}^m,
 \end{align}
 
 \item[{\bf F3)}] Order $N$ in $z$, order $M$ is $\mathcal{S}$:
 \begin{align}
    \formF(q^2,s) = \frac{1}{1 - \frac{q^2}{m_P^2}} \sum_{n=0}^{N} \sum_{m=0}^{M} A_{nm} z^n \mathcal{S}^m.
 \end{align} 
\end{itemize}
The first two families, {\bf F1} and {\bf F2}, cut the series at the combined $z$ and ${\cal S}$ order, while the third family {\bf F3} separately specifies the orders in $z$ and ${\cal S}$. In the limit of large $K$, $N$, $M$, all parametrizations become equal. \\

In the construction of $\chi^2$, we take into account the uncertainties in all $z$ and $s$ values by promoting these values to nuisance parameters, like
we did (for the $s$ values) in Ref.~\cite{Alexandrou:2017mpi}. The covariance matrix, which we estimate using single-elimination jackknife, is therefore a $3N_{\rm data}\times 3N_{\rm data}$ matrix, where $N_{\rm data}=48$ is the number of kinematic points. We added the systematic uncertainties associated with the choices of fit ranges in the matrix element fits and spectrum fits in quadrature to the diagonal elements of the covariance matrix. The uncertainties of the best-fit parameters are obtained from the Hessian of $\chi^2$ at the minimum.

%-----------------------------------------------------------------%
\subsection{The fit results}\label{subsec_V_fits}
%-----------------------------------------------------------------%
For each of the different families of parametrizations {\bf F1}-{\bf F3} we investigate several fits while keeping the power of $z$ below $3$, and power of ${\cal S}$ below $4$. We find that when the $z$-expansion goes to order $n=3$ or higher, the additional parameters are consistent with zero and no longer contribute to the description of the data; similarly, for the ${\cal S}$ expansion, at order $m=4$ the parameters become statistically consistent with zero. We drop all parametrizations yielding fit parameters with uncertainties larger than $100$ times their central values. We also remove parametrizations that lead to $\frac{\chi^2}{\rm dof} > 1.1$, which includes those that are of $0$-th order in the $z$-expansion. The list of models that we keep in our analysis, and their corresponding values of $\frac{\chi^2}{\rm dof}$, are given in Table \ref{tab:chi2}.

\begin{center}
\begin{figure}[htb]
  \includegraphics[width=\linewidth]{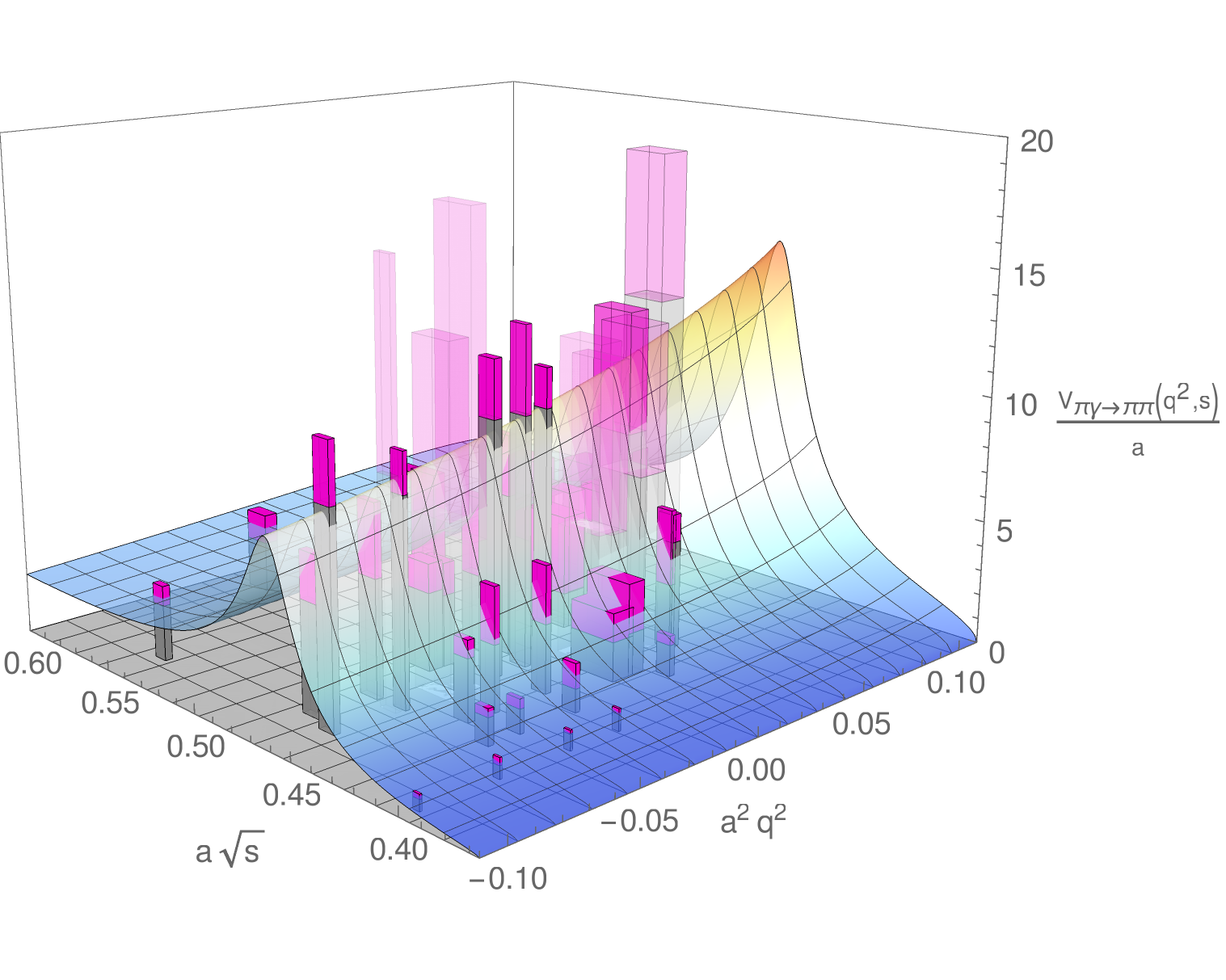}
  
  \caption{\label{fig:V_3d} Three-dimensional plot of the transition amplitude $\ampV$ (in lattice units) as a function of $\sqrt{s}$ and $q^2$. The lattice QCD results are shown as the vertical bars, where the widths and depths correspond to the uncertainties in $a\sqrt{s}$ and $a^2q^2$, and the magenta sections at the tops cover the range from $\ampV-\sigma_{\ampV}$ to $\ampV+\sigma_{\ampV}$. Data points with larger uncertainty are plotted with reduced opacity. The surface shows the central value of the nominal fit function (``BWII F1 K2''). }
\end{figure}
\end{center}
\begin{center}
\begin{figure}

  \vspace{10ex}

  \includegraphics[width=\linewidth]{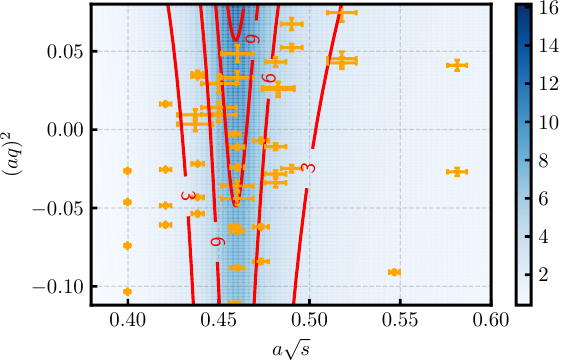}
  \caption{\label{fig:V_contour} Density plot of the fitted transition amplitude $\ampV$ (in lattice units, nominal parametrization ``BWII F1 K2'') in the $a\sqrt{s}$ and $(aq)^2$ plane. The locations of the discrete lattice QCD data points are indicated by the orange points with error bars.}
\end{figure}
\end{center}
\begin{center}
\begin{figure}
  \includegraphics[width=\linewidth]{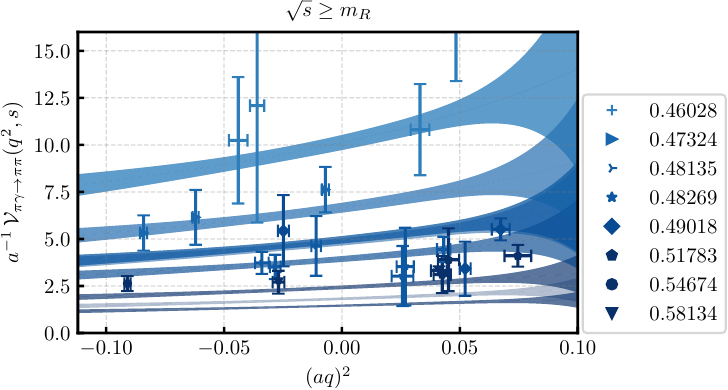}
  
  \vspace{2ex}
  
  \includegraphics[width=\linewidth]{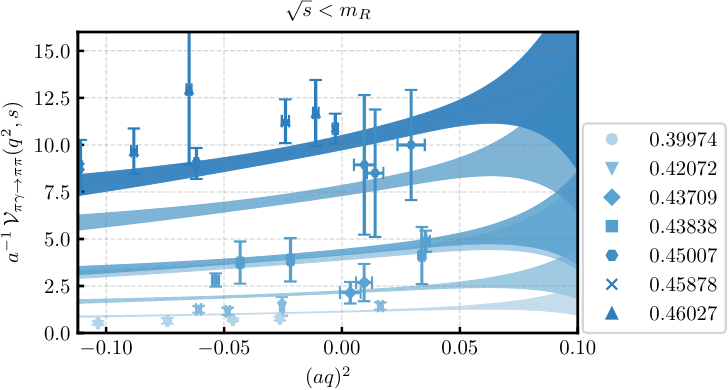}
  \caption{\label{fig:V_slice_s} The transition amplitude $\ampV$ (in lattice units, nominal parametrization ``BWII F1 K2''), sliced by value of invariant mass $\sqrt{s}$, as a function of $q^2$. The shaded bands correspond to the $1\sigma$ regions of the fitted parametrizations; their colors and brightness match the data points at the same $a\sqrt{s}$, as indicated by the symbols in the legend.}
\end{figure}
\end{center}
\begin{center}
\begin{figure}
  \includegraphics[width=0.8\linewidth]{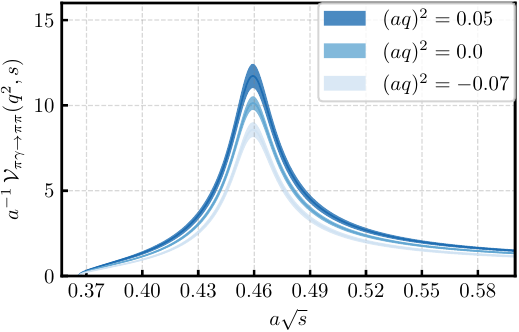}
  
  \includegraphics[width=0.8\linewidth]{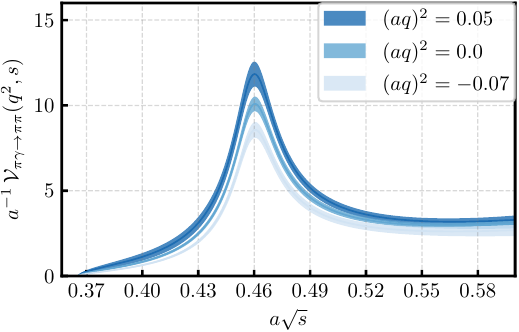}
  \caption{\label{fig:V_slice_q2} The transition amplitude $\ampV$ as a function of $\pi\pi$ invariant mass, for three different values of the $q^2$. The top panel corresponds to the nominal parametrization ``BWII F1 K2'', and the bottom panel corresponds to the parametrization ``BWI F1 K2''.}
\end{figure}
\end{center}
\begin{center}
\begin{figure}
  \includegraphics[width=\linewidth]{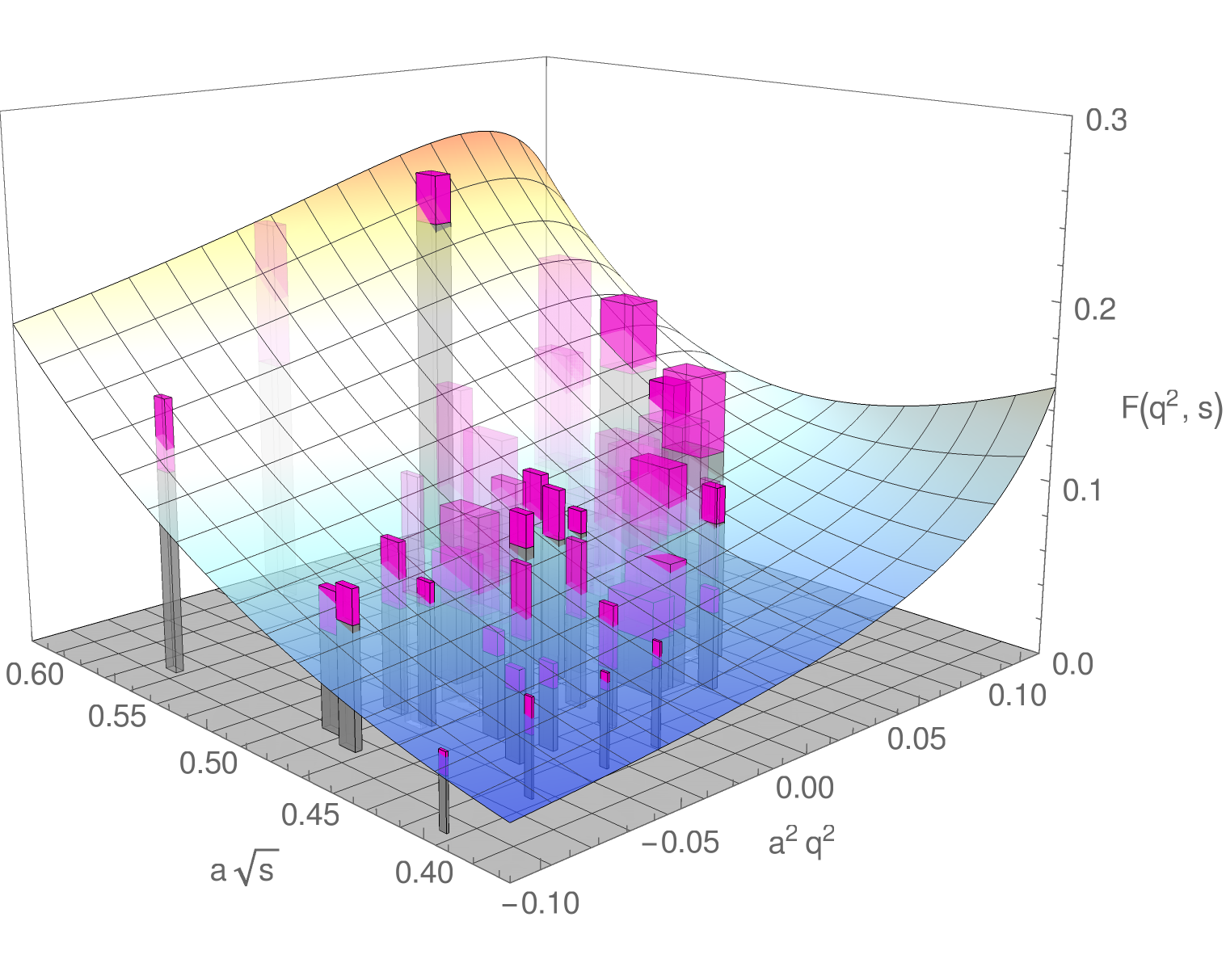}
  \caption{\label{fig:V_3d_sdep} Like Fig.~\protect\ref{fig:V_3d}, but for the function $\formF(q^2, s)$. The data points are divided by the central value of the Breit-Wigner factor (cf. Eq. \eqref{eq:ampV}) to represent the same quantity. }
\end{figure}
\end{center}
\begin{center}
\begin{figure}
  \includegraphics[width=\linewidth]{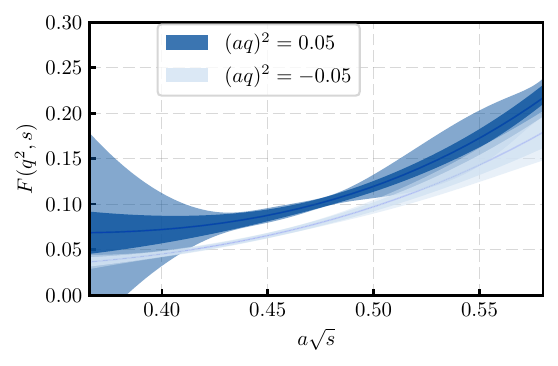}
  \includegraphics[width=\linewidth]{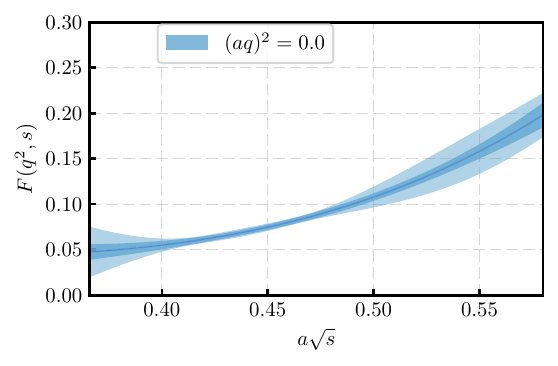}
  \caption{\label{fig:F_sdep} The form factor $\formF(q^2, s)$, as a function of $\pi\pi$ invariant mass, for two different nonzero values of $q^2$ (top) and for $q^2=0$ (bottom). Plotted is the central value of the nominal parametrization ``BWII F1 K2'' along with the two uncertainties: the inner (darker) shaded region represents the statistical and systematical uncertainties, and the outer (lighter) region includes also the parametrization uncertainty, estimated
  as the root-mean-square deviation of the central values obtained from the different parametrizations with respect to the nominal one.}
\end{figure}
\end{center}

\begin{table}[htb]
\begin{tabular}{|l|c|l|c|}
\hline
Parametrization & $\chi^2/{\rm dof}\:\:(\chi^2)$  \cr
\hline
BWI F1 K2       & 0.98 (41.25) \cr
BWI F1 K3       & 1.05 (39.99) \cr
BWI F2 N1 K2    & 0.97 (41.56) \cr 
BWI F2 N1 K3    & 0.99 (40.57) \cr
BWI F3 N1 M1  & 1.09 (47.90) \cr
BWI F3 N1 M2  & 0.99 (41.41) \cr 
BWI F3 N2 M2  & 1.04 (40.69) \cr
BWI F3 N2 M3  & 0.92 (33.23) \cr 
BWII F1 K2      & 1.07 (45.03) \cr
BWII F2 N1 K2   & 1.05 (45.14) \cr 
BWII F2 N1 K3   & 1.06 (43.53) \cr
BWII F3 N1 M2 & 1.07 (45.02) \cr 
BWII F3 N1 M3 & 1.07 (42.98) \cr
BWII F3 N2 M3 & 0.99 (35.68) \cr
\hline
\end{tabular}
\caption{\label{tab:chi2} List of parametrizations, and their values of $\chi^2/{\rm dof}$ and total $\chi^2$.}
\end{table}
We name the parametrizations according to the type of Breit-Wigner, family of truncation, and truncation limits.
The parametrizations that survive the cuts are consistent with each other within the uncertainties, and we choose ``BWII F1 K2''
as our nominal parametrization. All fit results are listed in Tables \ref{tab:fitparamsBWI} and \ref{tab:fitparamsBWII} in Appendix~\ref{app:MODEL_pars}.
The covariance matrix for the nominal parametrization is provided as part of the supplemental material \cite{supplementalmaterial}.

In Fig.~\ref{fig:V_3d} we present the fitted $\ampV$ combined with the data points in a three-dimensional plot as a function of $\sqrt{s}$ and $q^2$.
Figure \ref{fig:V_contour} instead shows a top-down view as a density plot, where the discrete values of $\sqrt{s}$ allowed by the finite volume for which
we have results appear as vertically aligned points.

The slices of the fitted amplitude at these discrete values of $\sqrt{s}$ are plotted as a function of $q^2$ in Fig.~\ref{fig:V_slice_s}, where the upper panel
shows the slices with $\sqrt{s}\ge m_R$ while the lower panel shows the slices with $\sqrt{s}< m_R$. We can see that the parametrization describes both the 
$\sqrt{s}$ and $q^2$ dependence of the data well.

Qualitatively, we can see two main features in $\ampV$: the amplitude is falling off as $q^2$ decreases, and shows the expected enhancement in $\sqrt{s}$ attributed to the $\rho$ resonance. The amplitude vanishes at the threshold $2 m_\pi$, then rises and falls steeply as the resonance region is crossed. This can also be seen in Fig.~\ref{fig:V_slice_q2}, where we plot $\ampV$ as a function of invariant mass for three different values of $q^2$. In this figure,
we show plots for both the nominal parametrization ``BWII F1 K2'' and for the parametrization ``BWI F1 K2'' that does not include the Blatt-Weisskopf barrier factor. At large $\sqrt{s}$, these parametrizations show some deviation. Nevertheless, for both parametrizations, the falloff of the amplitude at large $\sqrt{s}$ is slower than what would be expected for purely resonant behavior, indicating that the $\pi\gamma\to\pi\pi$ transition probability remains sizable even when the invariant mass is far above the resonance position. This is also reflected in Figs.~\ref{fig:V_3d_sdep} and \ref{fig:F_sdep}, where we plot the function $\formF(q^2, s)$ that does not contain the Breit-Wigner factor. The slow falloff of $\ampV$ as a function of $\sqrt{s}$
corresponds to growing $\formF$. The other parametrizations show the same behavior, confirming a nontrivial $s$-dependence of the function $\formF(q^2, s)$.

%

%-----------------------------------------------------------------%
\section{Observables}\label{sec_physics}
%-----------------------------------------------------------------%

\begin{center}
\begin{figure}[htb]
	\includegraphics[width=0.47\textwidth]{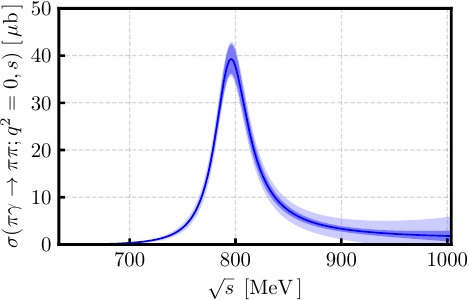}
	\caption{\label{fig:x_section} The $\pi^+\gamma\to\pi^+\pi^0$ photoproduction cross section as a function of $\pi^+\pi^0$ invariant mass, computed with the nominal parametrization ``BWII F1 K2'' of the amplitude, for our pion mass of $m_\pi\approx 320$ MeV. The inner (darker) shaded region indicates the statistical and systematic uncertainties, and the outer (lighter) shaded region also includes the parametrization uncertainty, estimated as explained in the caption of Fig.~\protect\ref{fig:F_sdep}.}
\end{figure}
\end{center}

As discussed in Sec.~\ref{sec_about_photproduction} we consider two main observable
quantities, both with a real photon ($q^2=0$): the $\pi^+\gamma\to\pi^+\pi^0$ cross section and the $\rho$ radiative decay width.
The $\pi^+\gamma\to\pi^+\pi^0$ cross section (\ref{eq:xsection}) evaluated with our nominal parametrization ``BWII F1 K2'' of $\ampV(s,q^2=0)$
is shown in Fig.~\ref{fig:x_section}. Note that we evaluated Eq.~(\ref{eq:xsection}) using the heavier-than-physical pion mass of this ensemble,
$m_\pi\approx 320$ MeV. Because the $\rho$ resonance is narrower than in nature, the peak value of the cross section is higher \cite{Briceno:2016kkp}.

\begin{center}
\begin{figure}[htb]
  \includegraphics[width=0.49\textwidth]{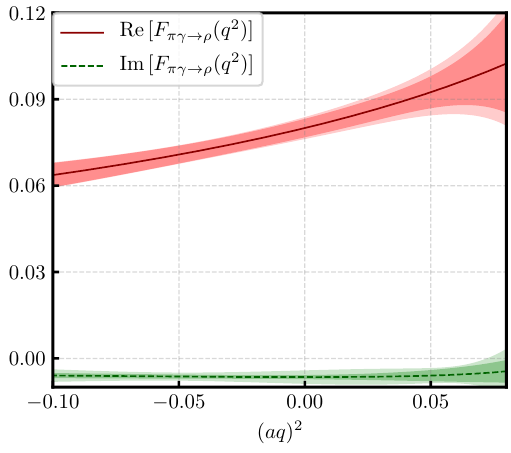}
  \caption{\label{fig:form_factor} The real and imaginary parts of the resonant form factor $F_{\pi\gamma\to\rho}(q^2)$ obtained by analytically continuing   the nominal parametrization ``BWII F1 K2'' of the $\pi\gamma\to\pi\pi$ amplitude to the $\rho$ resonance pole. The inner (darker) shaded region indicates the statistical and systematic uncertainties, and the outer (lighter) shaded region also includes the parametrization uncertainty, estimated as explained in the caption of Fig.~\protect\ref{fig:F_sdep}.}
\end{figure}
\end{center}

\begin{center}
\begin{figure}[htb]
  \includegraphics[width=0.47\textwidth]{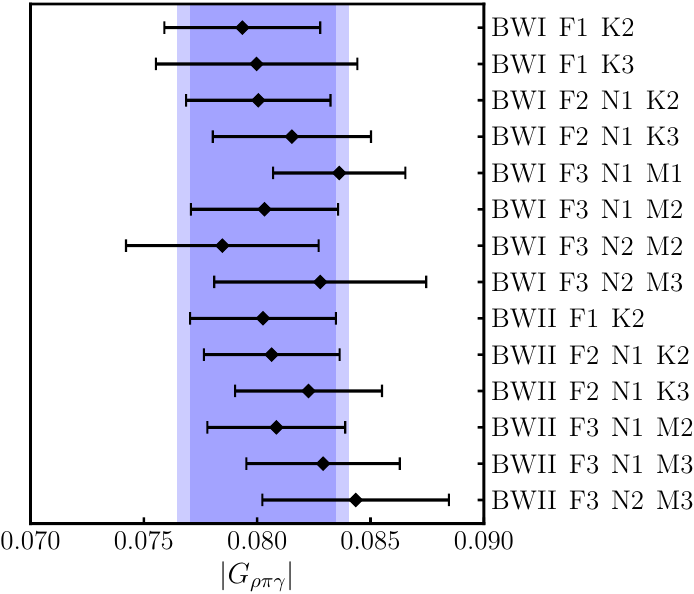}
  \caption{\label{fig:photocoupling} The $\rho$ meson photocoupling determined from the fourteen different parametrizations of the $\pi\gamma \to \pi\pi$ amplitudes. The bands indicate the value and uncertainties obtained from the nominal parametrization ``BWII F1 K2'', where the outer (lighter) band includes (added in quadrature) the root-mean-square deviation between all parametrizations and the chosen one.}
\end{figure}
\end{center}

To determine the $\rho$ radiative decay width, $\Gamma(\rho \to \pi \gamma)$, we must first determine the photocoupling $G_{\rho\pi\gamma}$,
which requires us to analytically continue the transition amplitude $\ampV$ to the pole position. The resulting resonant form factor $F_{\pi\gamma\to\rho}(q^2)$,
defined in Eq.~\eqref{eq:Fresonant}, is presented in Fig.~\ref{fig:form_factor}. We find that the imaginary part of the resonant form factor is consistent with zero,
and the real part slowly rises as a function of $q^2$. The resonant form factor at $q^2=0$ is equal to the photocoupling, $G_{\rho\pi\gamma} = F_{\pi\gamma\to\rho}(0)$.
Our results for $G_{\rho\pi\gamma}$, now for all fourteen amplitude parametrizations that gave good fits, are shown in Fig.~\ref{fig:photocoupling}. 
We find that the photocouplings extracted from the different parametrizations are consistent with each other. Nevertheless, we estimate a systematic uncertainty associated with the choice of parametrization as
\begin{align}
\sqrt{\sum_{i=1}^N\frac{ (x_i - x_{\rm chosen})^2}{N-1}}, \label{eq:sigmasystparam}
\end{align}
where $x_i$ is the photocoupling determined from the $i$-th parametrizations,
$N=14$ is the number of different parametrizations, and $x_{\rm chosen}$ is the value obtained from the nominal parametrization, ``BWII F1 K2''.
Our final result for the photocoupling is
\begin{align}
|G_{\rho\pi\gamma}| = 0.0802(32)(20),
\end{align}
where the first uncertainty includes the statistical uncertainty and the systematic uncertainty from the two-point and three-point function fits,
while the second uncertainty is our estimate \eqref{eq:sigmasystparam} of the parametrization dependence.

The kinematic factors in Eq.~\eqref{eq:radwidth} lead to a strong pion-mass dependence of the $\rho$ radiative decay width. We can calculate
the decay width for the physical pion mass under the assumption that the pion-mass dependence of the photocoupling is negligible. This gives
\begin{align}
\Gamma(\rho\to\pi\gamma) \,=\, 168(13)(8)\, {\rm keV},
\end{align}
where we used $m_\rho=775$ MeV and $m_\pi=140$ MeV. For comparison, the experimental value of the $\rho^\pm$ radiative decay width is $68(7)$ keV \cite{Patrignani:2016xqp}.

%%%%%%%%%%%%%%%%%%%%%%%%%%%%%%%%%%%%%%%%%%%%%%%%%%%%%%%%%%%%%%%%%%%
\section{CONCLUSIONS}\label{sec_summary}
%%%%%%%%%%%%%%%%%%%%%%%%%%%%%%%%%%%%%%%%%%%%%%%%%%%%%%%%%%%%%%%%%%%

We have presented a $(2+1)$-flavor lattice QCD calculation of the $\pi \gamma \to \pi\pi$ process, where the $\pi\pi$ system has $I=1$ and $J^{PC}=1^{--}$.
The ensemble used has light-quark masses that correspond to a pion mass of approximately 320 MeV, while the strange-quark mass is approximately at its physical value. For the $\pi\pi$ system, we utilized the same moving frames and irreducible representations as in our previous study of $\pi\pi$ scattering \cite{Alexandrou:2017mpi}.
We determined the transition amplitude $\ampV(q^2,s)$ with few-percent uncertainty in a broad kinematic region around the $\rho$ pole in invariant mass $s$ and around zero momentum transfer $q^2$, using model-independent parametrizations based on a series expansion in the variables $z$ and $\mathcal{S}$, defined in Eqs.~\eqref{eq:z} and \eqref{eq:calS}. The results obtained from several different truncations of the series are consistent with each other. We observe the expected enhancement
of the amplitude associated with the $\rho$ resonance, but find that for large $\sqrt{s}$ the amplitude falls off slower than expected for purely resonant behavior.
In our analysis, we compared two different Breit-Wigner parametrizations of the $\pi\pi$ scattering phase shift (with and without a Blatt-Weisskopf barrier factor).
These parametrizations yield consistent results for $\ampV(q^2,s)$ in most of the kinematic range, but differ for large $\sqrt{s}$.

By analytically continuing $\ampV(q^2,s)$ to the $\rho$ pole, we also determined the $\pi\gamma \to \rho$ resonant form factor and the $\rho$ photocoupling.
All truncations of the series used for $\ampV(q^2,s)$, and both Breit-Wigner functions, lead to consistent results for the photocoupling, as can be seen in
Fig.~\ref{fig:photocoupling}. Our final result for this coupling is $|G_{\rho\pi\gamma}| = 0.0802(32)(20)$, which is approximately $1.6$ times the value extracted from the measured $\rho^\pm$ radiative decay width \cite{Patrignani:2016xqp} using Eq.~\eqref{eq:radwidth}, $|G_{\rho\pi\gamma}|_{\rm exp}=0.0508 (26)$. A significant difference between the lattice result and the experimental value can be expected, given that we did not perform an extrapolation to zero lattice spacing and physical pion mass, and we did not estimate the resulting systematic errors.
Most of the past lattice studies of this quantity \cite{Woloshyn:1986pk,Crisafulli:1991pn,Owen:2015fra,Shultz:2015pfa} were performed in the single-hadron approach, in which the coupling of the $\rho$ to the $\pi\pi$ system is not taken into account. The authors of Refs.~\cite{Briceno:2015dca,Briceno:2016kkp} used the multi-hadron approach at a pion mass of approximately 400 MeV and obtained a value of $|G_{\rho\pi\gamma}|$ around $0.12$, as can be seen in Fig.~12 of \cite{Briceno:2016kkp}.

Future calculations at lower pion masses, larger volumes, and additional values of the lattice spacing are needed to extrapolate to the physical point.
One aspect that also requires more attention is the residual contamination from higher excited states in the ratios used to determine the matrix elements
from the correlation functions. Better control over this contamination can be achieved by using more than three source-sink separations and employing more advanced analysis methods \cite{Bulava:2011yz}.

The lattice methods used here to compute a $1\to 2$ transition are also applicable to many other processes of interest in nuclear and high-energy
physics. An important example is the rare decay $B\to K^*(\to K \pi)\ell^+ \ell^-$ \cite{Horgan:2013hoa,Horgan:2013pva}; new lattice calculations of the $B\to K^*$
form factors that take into account the strong decay of the $K^*$ are needed.

%%%%%%%%%%%%%%%%%%%%%%%%%%%%%%%%%%%%%%%%%%%%%%%%%%%%%%%%%%%%%
\acknowledgments

We are grateful to Kostas Orginos for providing the gauge-field ensemble, which was generated using resources provided by XSEDE (supported by National Science  Foundation Grant No.~ACI-1053575). We thank R.~A.~Brice\~no, M.~Hansen, M.~Hoferichter, B.~Kubis, C.~B.~Lang, and M.~Niehus for valuable discussions.
SM and GR were supported in part by National Science Foundation Grant No.~PHY-1520996; SM, GR, and LL were also supported in part by the U.S.~Department of Energy Office of High Energy Physics under Grant No.~DE-{}SC0009913. SM and SS futher acknowledge support by the RHIC Physics Fellow Program of the RIKEN BNL Research Center. JN and AP were supported in part by the U.S. Department of Energy Office of Nuclear Physics under Grant Nos.~DE{}-SC-0{}011090 and DE-{}FC02-06ER41444. We acknowledge funding from the  European Union's Horizon 2020 research and innovation programme under the Marie Sklodowska-Curie grant agreement No 642069. S.~P.~is a Marie Sklodowska-Curie  fellow supported by the HPC-LEAP joint doctorate program. This research used resources of the National Energy Research Scientific Computing Center (NERSC), a U.S. Department of Energy Office of Science User Facility operated under Contract No. DE-{}AC02-05CH11231. The computations were performed using the Qlua software suite \cite{QLUA}.

%%%%%%%%%%%%%%%%%%%%%%%%%%%%%%%%%%%%%%%%%%%%%%%%%%%%%%%%%%%%%

%%%%%%%%%%%%%%%%%%%%%%%%%%%%%%%%%%%%%%%%%%%%%%%%%%%%%%%%%%%%%
\appendix

\FloatBarrier
%%%%%%%%%%%%%%%%%%%%%%%%%%%%%%%%%%%%%%%%%%%%%%%%%%%%%%%%%%%%%%%%%%%%%%%%%%%%%%%%%%%%%%%%%%%%%%%%%%%%%%%%%%%%%%%
\section{MATRIX ELEMENT FITS}\label{app:ME_fits}
%%%%%%%%%%%%%%%%%%%%%%%%%%%%%%%%%%%%%%%%%%%%%%%%%%%%%%%%%%%%%%%%%%%%%%%%%%%%%%%%%%%%%%%%%%%%%%%%%%%%%%%%%%%%%%%
\FloatBarrier

\begin{figure*}
\includegraphics[width=0.98\textwidth]{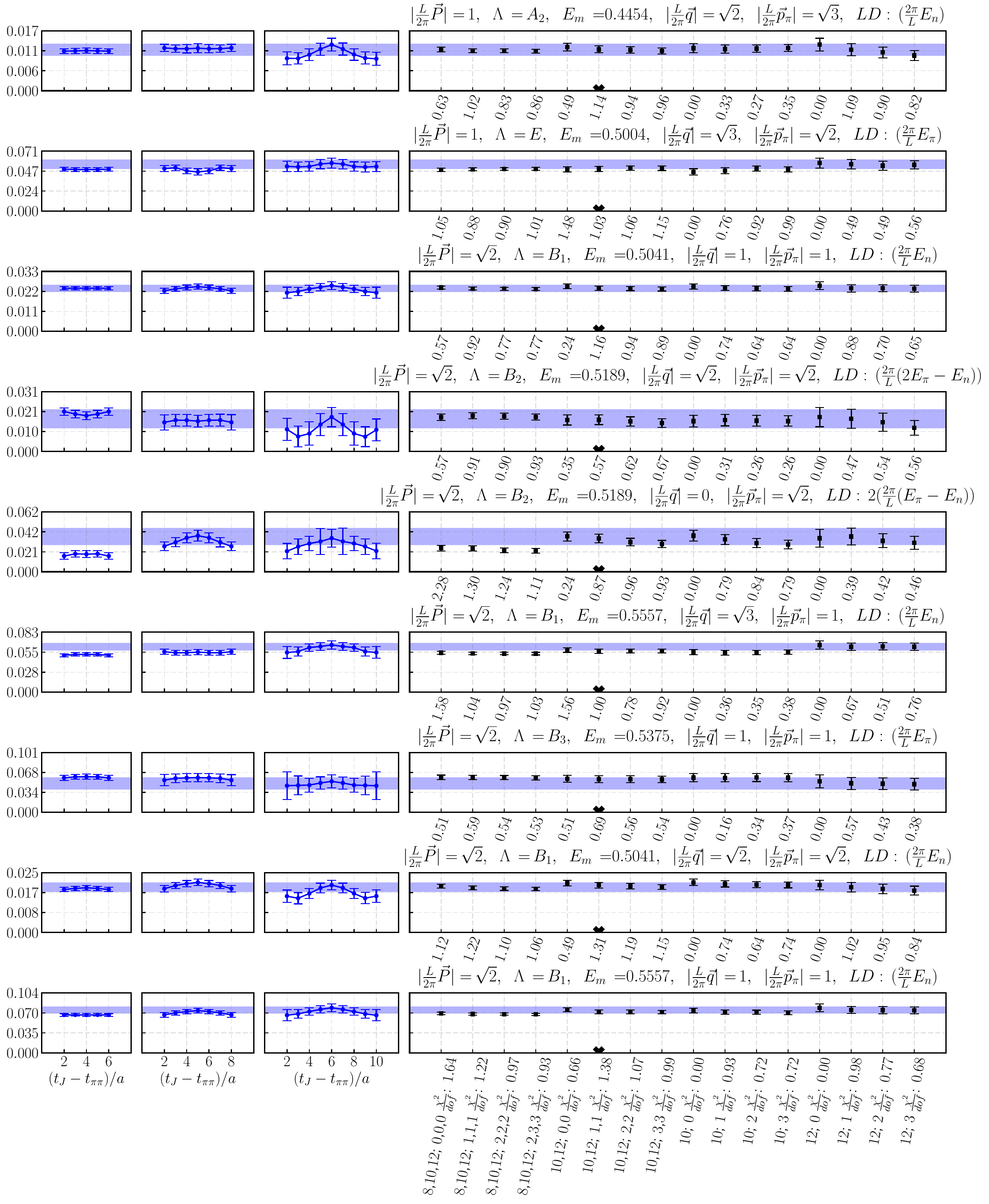}
\caption{\label{fig:ev_i1} As in Fig.~\protect\ref{fig:example_ME_fit}, for additional kinematic points.}
\end{figure*}

\begin{figure*}
\includegraphics[width=0.98\textwidth]{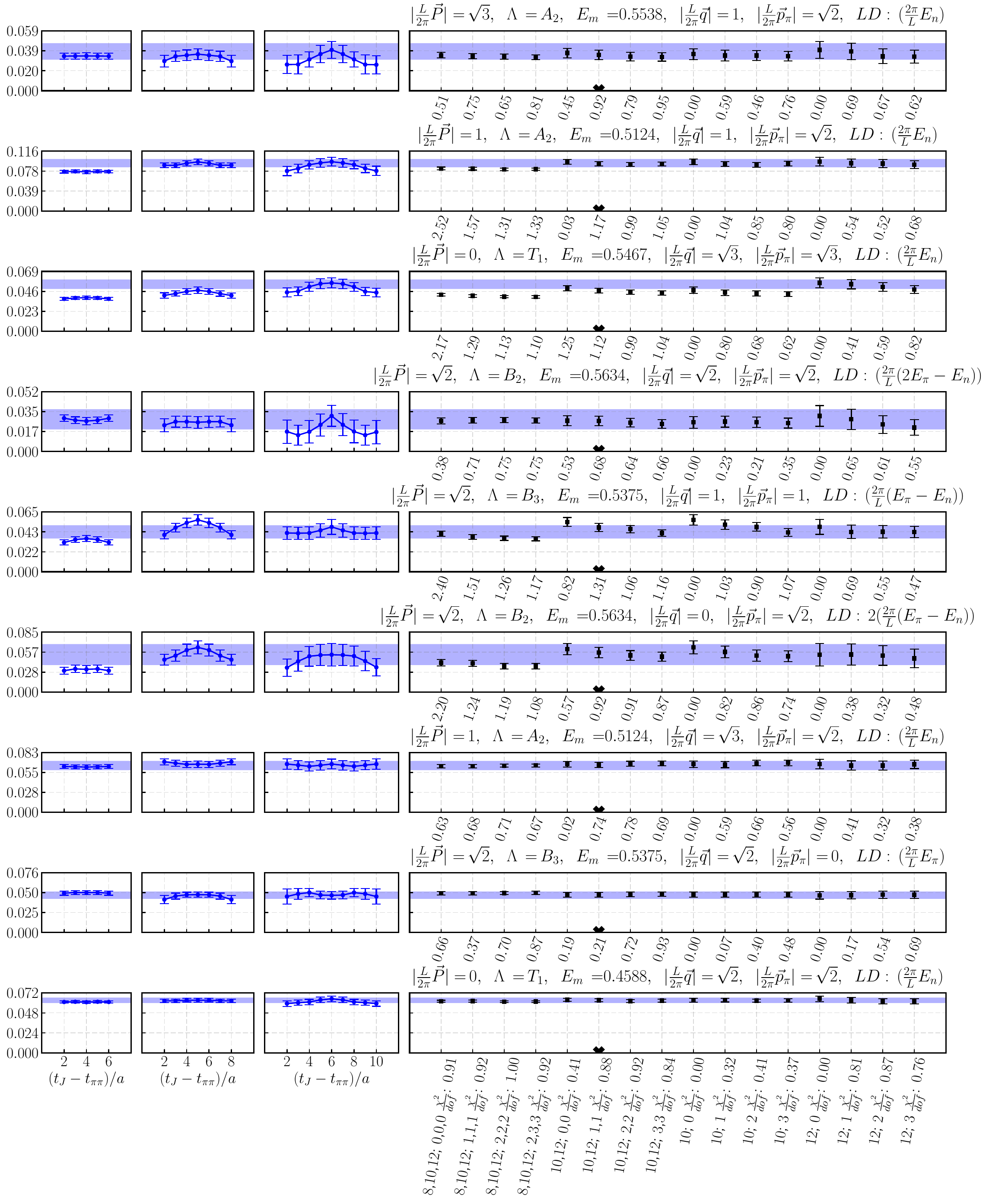}
\caption{\label{fig:ev_i2} As in Fig.~\protect\ref{fig:example_ME_fit}, for additional kinematic points.}
\end{figure*}

\begin{figure*}
\includegraphics[width=0.98\textwidth]{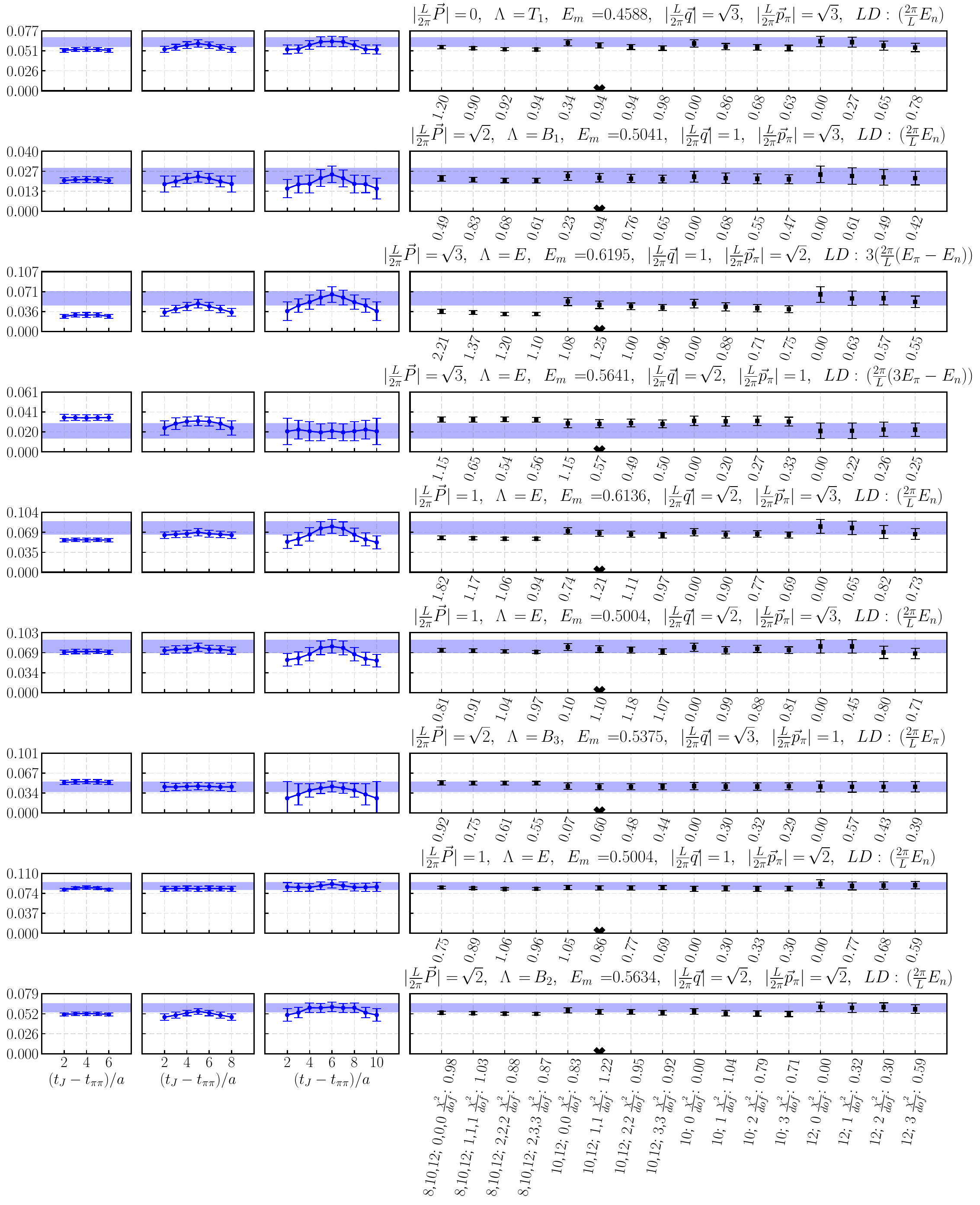}
\caption{\label{fig:ev_i3} As in Fig.~\protect\ref{fig:example_ME_fit}, for additional kinematic points.}
\end{figure*}

\begin{figure*}
\includegraphics[width=0.98\textwidth]{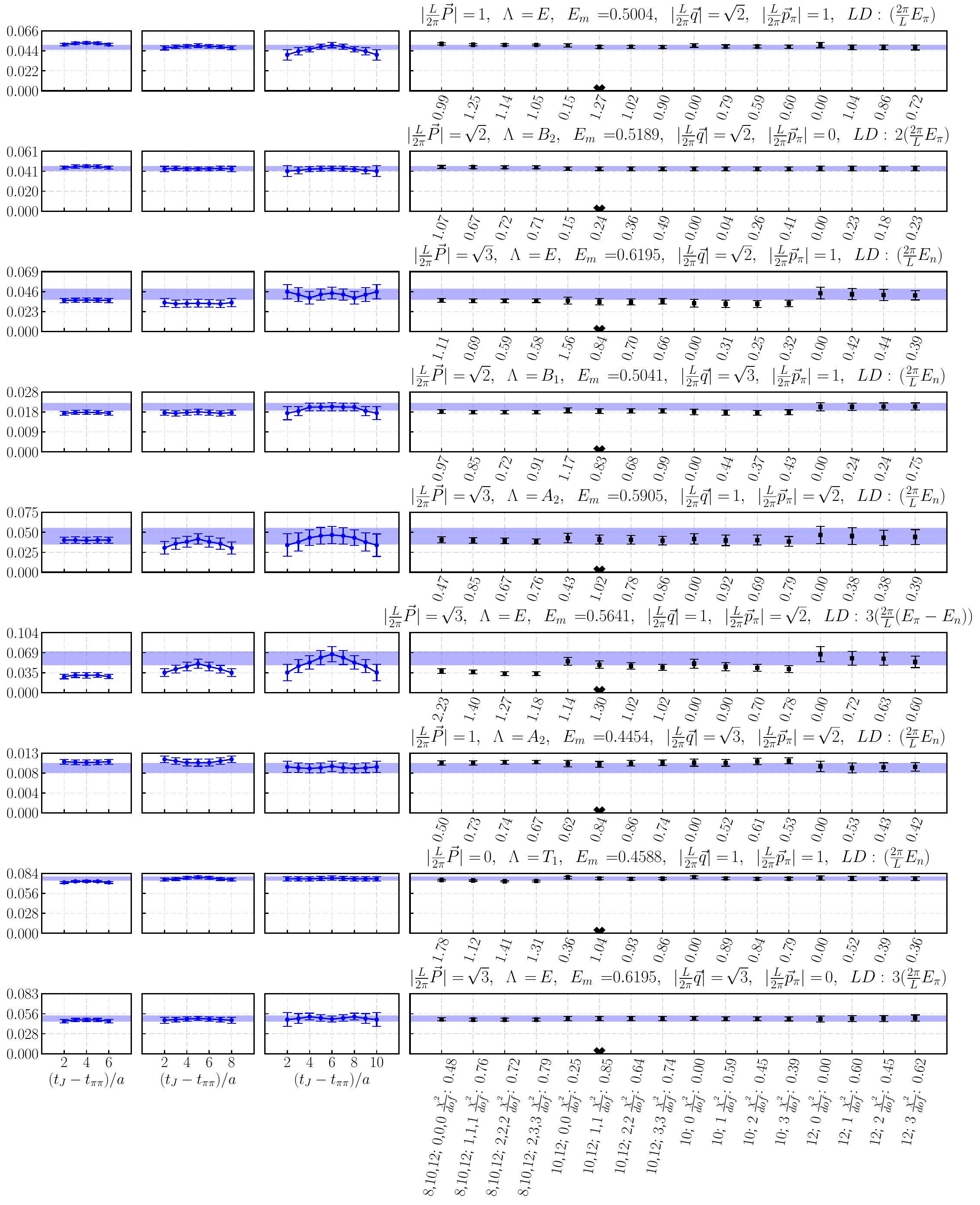}
\caption{\label{fig:ev_i4} As in Fig.~\protect\ref{fig:example_ME_fit}, for additional kinematic points.}
\end{figure*}

\begin{figure*}
\includegraphics[width=0.98\textwidth]{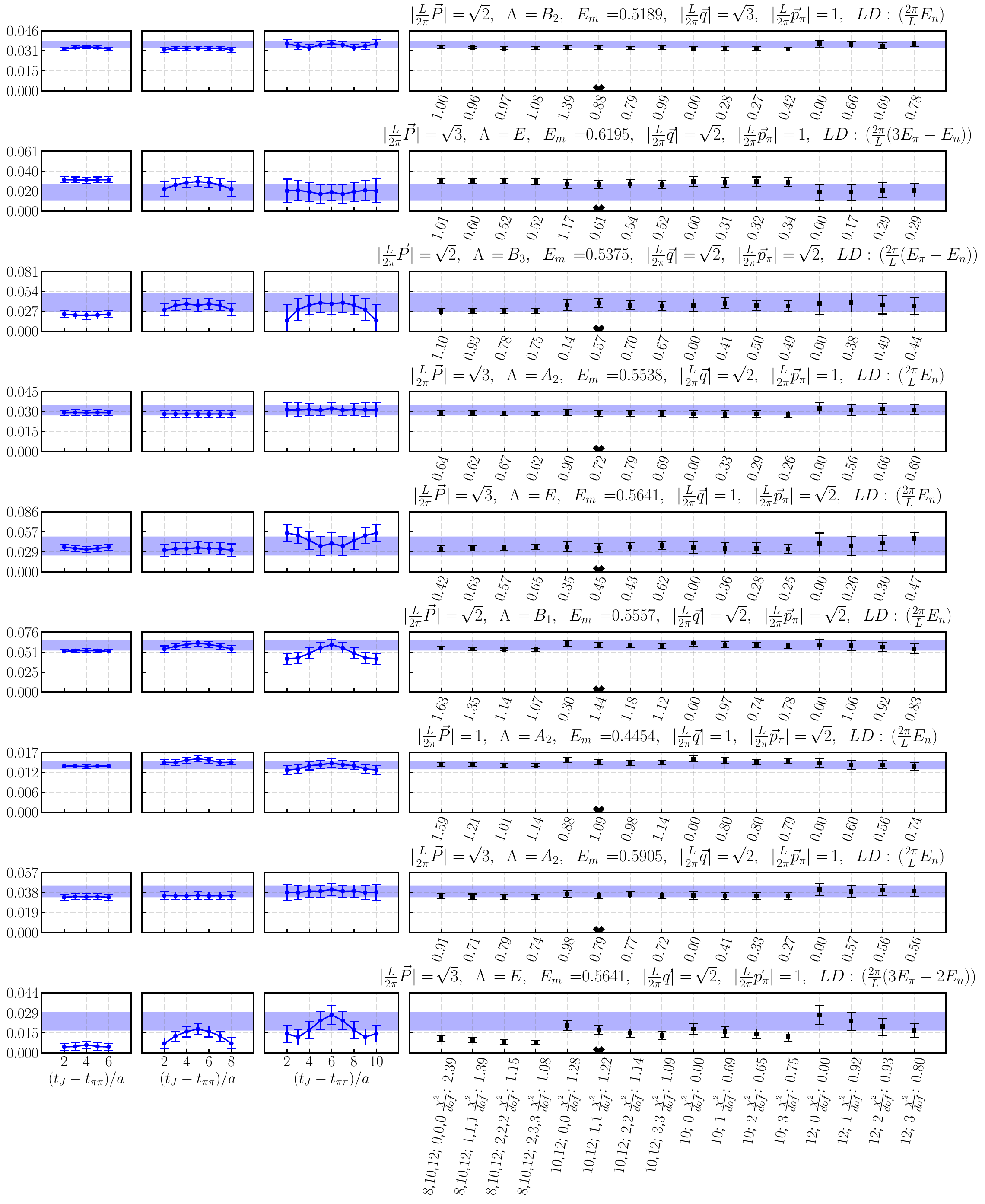}
\caption{\label{fig:ev_i5} As in Fig.~\protect\ref{fig:example_ME_fit}, for additional kinematic points.}
\end{figure*}

\begin{figure*}
\includegraphics[width=0.98\textwidth]{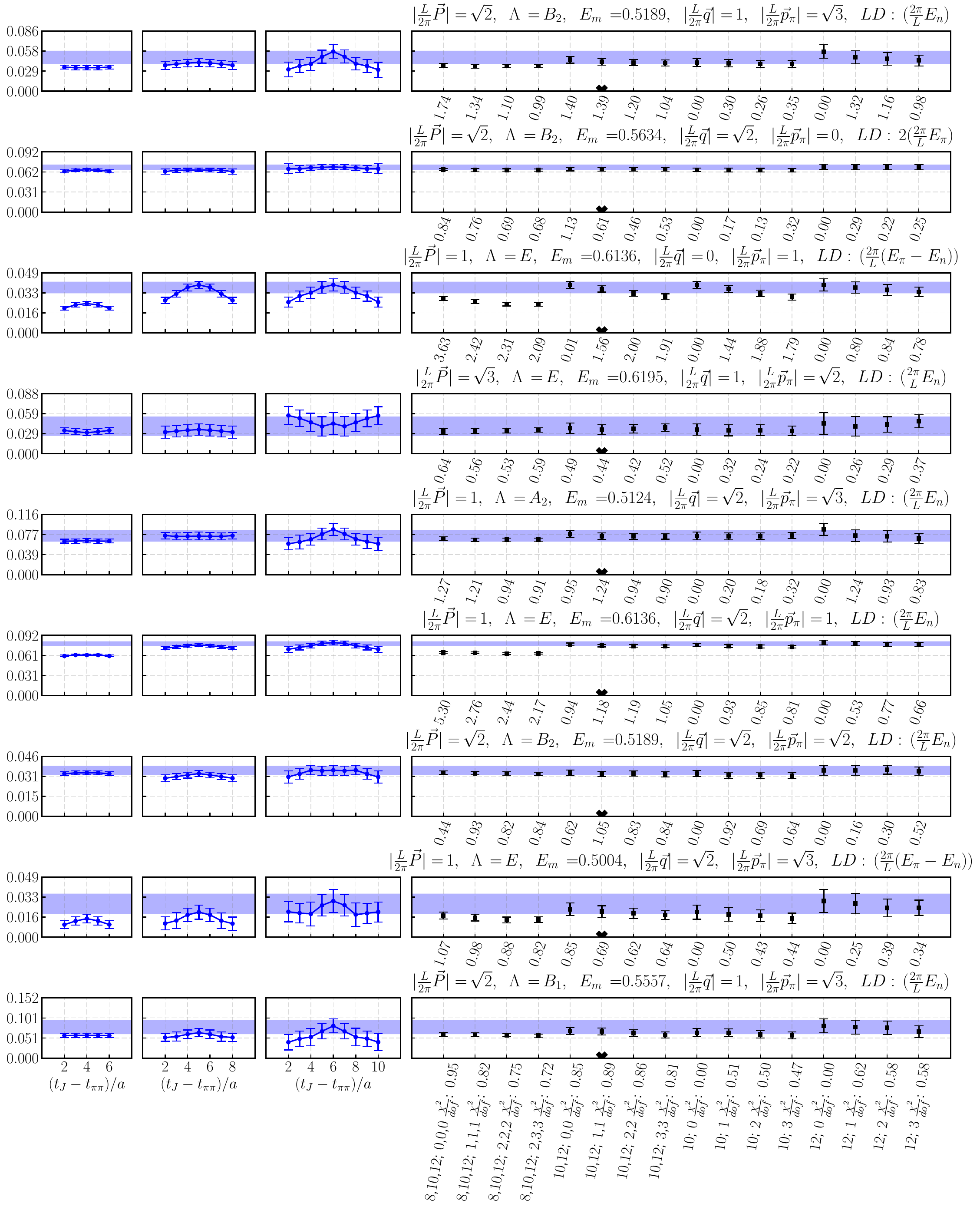}
\caption{\label{fig:ev_i6} As in Fig.~\protect\ref{fig:example_ME_fit}, for additional kinematic points.}
\end{figure*}

Figures \ref{fig:ev_i1}--\ref{fig:ev_i6} show the ratios used to extract the finite-volume matrix elements
and the fit results for multiple different fit ranges, for additional kinematic points that were omitted
in Fig.~\ref{fig:example_ME_fit}. Tables \ref{tab:MEpt1} and \ref{tab:MEpt2} give the values of both the finite-volume and
infinite-volume matrix matrix elements for all kinematic points.

\begin{table*}
\centering
\begin{tabular}{|c|c|c|c|c|c|c|c|c|}
\hline
$|\frac{L}{2\pi}\vec{P}|$ & $\Lambda$ & $|\frac{L}{2\pi}\vec{p}_{\pi}|$ & LD & $\sqrt{s_n^{\vec{P},\,\Lambda}}$ & $(q^2)_n^{\vec{P},\,\Lambda}$ & $ME_{FV}$ & $ME_{IV}^{BWI}$ & $ME_{IV}^{BWII}$ \cr
\hline
$0$ & $T_1$ & $1$          & $(\frac{2 \pi}{L} E_n)$ & $0.4588(29)$ & $-0.0029(11)$ & $0.0767(45)$ & $10.81(96)$ & $10.70(79)$ \cr
$0$ & $T_1$ & $\sqrt{ 2 }$ & $(\frac{2 \pi}{L} E_n)$ & $0.4588(29)$ & $-0.06173(81)$ & $0.0636(50)$ & $8.97(93)$ & $8.88(81)$ \cr
$0$ & $T_1$ & $\sqrt{ 3 }$ & $(\frac{2 \pi}{L} E_n)$ & $0.4588(29)$ & $-0.11087(64)$ & $0.0624(95)$ & $8.8(1.5)$ & $8.7(1.4)$ \cr
$0$ & $T_1$ & $\sqrt{ 3 }$ & $(\frac{2 \pi}{L} E_n)$ & $0.5467(28)$ & $-0.0910(13)$ & $0.0545(83)$ & $3.06(46)$ & $3.08(47)$ \cr
$1$ & $A_2$ & $1$          & $(\frac{2 \pi}{L} E_n)$ & $0.3997(14)$ & $-0.04630(41)$ & $0.0125(12)$ & $0.657(68)$ & $0.669(69)$ \cr
$1$ & $A_2$ & $\sqrt{ 2 }$ & $(\frac{2 \pi}{L} E_n)$ & $0.3997(14)$ & $-0.10343(39)$ & $0.0095(16)$ & $0.499(86)$ & $0.509(88)$ \cr
$1$ & $A_2$ & $\sqrt{ 2 }$ & $(\frac{2 \pi}{L} E_n)$ & $0.3997(14)$ & $-0.02632(39)$ & $0.0136(19)$ & $0.72(10)$ & $0.73(10)$ \cr
$1$ & $A_2$ & $\sqrt{ 2 }$ & $(\frac{2 \pi}{L} E_n)$ & $0.4732(42)$ & $-0.0841(15)$ & $0.0649(99)$ & $5.9(1.1)$ & $5.8(1.0)$ \cr
$1$ & $A_2$ & $\sqrt{ 2 }$ & $(\frac{2 \pi}{L} E_n)$ & $0.4732(42)$ & $-0.0070(15)$ & $0.093(12)$ & $8.5(1.4)$ & $8.4(1.3)$ \cr
$1$ & $A_2$ & $\sqrt{ 3 }$ & $(\frac{2 \pi}{L} E_n)$ & $0.3997(14)$ & $-0.07400(41)$ & $0.0116(26)$ & $0.61(14)$ & $0.62(14)$ \cr
$1$ & $A_2$ & $\sqrt{ 3 }$ & $(\frac{2 \pi}{L} E_n)$ & $0.4732(42)$ & $-0.0620(13)$ & $0.075(17)$ & $6.8(1.6)$ & $6.8(1.6)$ \cr
$1$ & $E$ & $1$          & $(\frac{2 \pi}{L} E_\pi)$ & $0.4603(37)$ & $-0.0240(16)$ & $0.0479(42)$ & $6.62(74)$ & $6.52(67)$ \cr
$1$ & $E$ & $1$          & $(\frac{2 \pi}{L} E_\pi)$ & $0.5813(53)$ & $0.0411(33)$ & $0.0481(43)$ & $2.22(20)$ & $2.23(20)$ \cr
$1$ & $E$ & $1$          & $(\frac{2 \pi}{L} (E_\pi - E_n))$ & $0.5813(53)$ & $0.1182(33)$ & $0.0371(72)$ & $1.71(33)$ & $1.72(33)$ \cr
$1$ & $E$ & $1$          & $(\frac{2 \pi}{L} E_n)$ & $0.5813(53)$ & $0.0411(33)$ & $0.0790(51)$ & $3.64(24)$ & $3.67(25)$ \cr
$1$ & $E$ & $\sqrt{ 2 }$ & $(\frac{2 \pi}{L} (E_\pi - E_n))$ & $0.4603(37)$ & $-0.0111(12)$ & $0.0323(74)$ & $4.5(1.1)$ & $4.4(1.0)$ \cr
$1$ & $E$ & $\sqrt{ 2 }$ & $(\frac{2 \pi}{L} E_n)$ & $0.4603(37)$ & $-0.0882(12)$ & $0.0697(84)$ & $9.6(1.4)$ & $9.5(1.3)$ \cr
$1$ & $E$ & $\sqrt{ 2 }$ & $(\frac{2 \pi}{L} E_\pi)$ & $0.4603(37)$ & $-0.0882(12)$ & $0.0554(84)$ & $7.7(1.3)$ & $7.5(1.2)$ \cr
$1$ & $E$ & $\sqrt{ 2 }$ & $(\frac{2 \pi}{L} E_n)$ & $0.4603(37)$ & $-0.0111(12)$ & $0.087(11)$ & $12.1(1.7)$ & $11.9(1.6)$ \cr
$1$ & $E$ & $\sqrt{ 3 }$ & $(\frac{2 \pi}{L} E_n)$ & $0.5813(53)$ & $-0.0269(25)$ & $0.077(17)$ & $3.53(79)$ & $3.56(80)$ \cr
$1$ & $E$ & $\sqrt{ 3 }$ & $(\frac{2 \pi}{L} E_n)$ & $0.4603(37)$ & $-0.0648(11)$ & $0.079(17)$ & $11.0(2.5)$ & $10.8(2.4)$ \cr
$1$ & $E$ & $\sqrt{ 3 }$ & $(\frac{2 \pi}{L} (E_\pi - E_n))$ & $0.4603(37)$ & $-0.0648(11)$ & $0.028(13)$ & $3.8(1.8)$ & $3.8(1.7)$ \cr
$\sqrt{ 2 }$ & $B_1$ & $1$          & $(\frac{2 \pi}{L} E_n)$ & $0.4207(30)$ & $-0.0608(11)$ & $0.0209(27)$ & $1.34(18)$ & $1.36(18)$ \cr
$\sqrt{ 2 }$ & $B_1$ & $1$          & $(\frac{2 \pi}{L} E_n)$ & $0.4207(30)$ & $0.0163(11)$ & $0.0238(31)$ & $1.53(20)$ & $1.55(21)$ \cr
$\sqrt{ 2 }$ & $B_1$ & $1$          & $(\frac{2 \pi}{L} E_n)$ & $0.4814(57)$ & $-0.0339(28)$ & $0.0628(80)$ & $4.45(72)$ & $4.45(70)$ \cr
$\sqrt{ 2 }$ & $B_1$ & $1$          & $(\frac{2 \pi}{L} E_n)$ & $0.4814(57)$ & $0.0432(28)$ & $0.0748(94)$ & $5.30(83)$ & $5.30(81)$ \cr
$\sqrt{ 2 }$ & $B_1$ & $\sqrt{ 2 }$ & $(\frac{2 \pi}{L} E_n)$ & $0.4207(30)$ & $-0.04843(86)$ & $0.0191(31)$ & $1.23(20)$ & $1.25(21)$ \cr
$\sqrt{ 2 }$ & $B_1$ & $\sqrt{ 2 }$ & $(\frac{2 \pi}{L} E_n)$ & $0.4814(57)$ & $-0.0283(23)$ & $0.0591(97)$ & $4.19(80)$ & $4.19(78)$ \cr
$\sqrt{ 2 }$ & $B_1$ & $\sqrt{ 3 }$ & $(\frac{2 \pi}{L} E_n)$ & $0.4207(30)$ & $-0.02545(91)$ & $0.0235(82)$ & $1.51(53)$ & $1.53(54)$ \cr
$\sqrt{ 2 }$ & $B_1$ & $\sqrt{ 3 }$ & $(\frac{2 \pi}{L} E_n)$ & $0.4814(57)$ & $-0.0110(20)$ & $0.078(26)$ & $5.5(1.9)$ & $5.5(1.9)$ \cr
$\sqrt{ 2 }$ & $B_2$ & $0$          & $2(\frac{2 \pi}{L} E_\pi)$ & $0.4384(33)$ & $0.0355(18)$ & $0.0433(40)$ & $3.77(44)$ & $3.84(44)$ \cr
$\sqrt{ 2 }$ & $B_2$ & $0$          & $2(\frac{2 \pi}{L} E_\pi)$ & $0.4902(58)$ & $0.0673(38)$ & $0.0689(62)$ & $4.32(47)$ & $4.34(46)$ \cr
$\sqrt{ 2 }$ & $B_2$ & $1$          & $(\frac{2 \pi}{L} E_n)$ & $0.4384(33)$ & $-0.0536(13)$ & $0.0354(38)$ & $3.08(39)$ & $3.14(39)$ \cr
$\sqrt{ 2 }$ & $B_2$ & $\sqrt{ 2 }$ & $(\frac{2 \pi}{L} E_n)$ & $0.4384(33)$ & $-0.0432(10)$ & $0.0355(56)$ & $3.09(52)$ & $3.15(53)$ \cr
$\sqrt{ 2 }$ & $B_2$ & $\sqrt{ 2 }$ & $(\frac{2 \pi}{L} (2E_\pi - E_n))$ & $0.4384(33)$ & $-0.0432(10)$ & $0.0170(74)$ & $1.48(65)$ & $1.51(67)$ \cr
$\sqrt{ 2 }$ & $B_2$ & $\sqrt{ 2 }$ & $2(\frac{2 \pi}{L} (E_\pi - E_n))$ & $0.4384(33)$ & $0.0339(10)$ & $0.037(13)$ & $3.2(1.2)$ & $3.3(1.2)$ \cr
$\sqrt{ 2 }$ & $B_2$ & $\sqrt{ 2 }$ & $(\frac{2 \pi}{L} E_n)$ & $0.4902(58)$ & $-0.0248(23)$ & $0.0603(91)$ & $3.78(62)$ & $3.80(62)$ \cr
$\sqrt{ 2 }$ & $B_2$ & $\sqrt{ 2 }$ & $(\frac{2 \pi}{L} (2E_\pi - E_n))$ & $0.4902(58)$ & $-0.0248(23)$ & $0.028(13)$ & $1.75(83)$ & $1.76(83)$ \cr
$\sqrt{ 2 }$ & $B_2$ & $\sqrt{ 2 }$ & $2(\frac{2 \pi}{L} (E_\pi - E_n))$ & $0.4902(58)$ & $0.0523(23)$ & $0.053(22)$ & $3.3(1.4)$ & $3.4(1.4)$ \cr
$\sqrt{ 2 }$ & $B_2$ & $\sqrt{ 3 }$ & $(\frac{2 \pi}{L} E_n)$ & $0.4384(33)$ & $-0.0218(11)$ & $0.049(14)$ & $4.3(1.3)$ & $4.3(1.3)$ \cr
\hline
\end{tabular}
\caption{\label{tab:MEpt1}Lattice results for the matrix elements (continued in Table \protect\ref{tab:MEpt2}), in lattice units. The quantity denoted as $LD$ is the kinematic factor appearing next to $2i \ampV/m_\pi$ in Eq.~(\ref{eq:LLdecomp}). Here, $ME_{FV}$ denotes the finite-volume matrix elements $|\langle \pi,\vecp | J_\mu(0,\vecQ)|n,\vecP,\,\Lambda, r \rangle_{FV}|$, after averaging over equivalent momentum directions and irrep rows $r$. The corresponding infinite-volume matrix elements,
with Lellouch-L\"uscher factors computed for the two different Breit-Wigner models, are denoted as $ME_{IV}^{BWI}$ and $ME_{IV}^{BWII}$. The systematic uncertainties
from the fits to the ratios (cf.~Sec.~\ref{sec_matrix_elements}) and from the spectrum fits (cf.~Ref.~\cite{Alexandrou:2017mpi}) have been added to the statistical uncertainties in quadrature.}
\end{table*}

\begin{table*}
\centering
\begin{tabular}{|c|c|c|c|c|c|c|c|c|}
\hline
$|\frac{L}{2\pi}\vec{P}|$ & $\Lambda$ & $|\frac{L}{2\pi}\vec{p}_{\pi}|$ & LD & $\sqrt{s_n^{\vec{P},\,\Lambda}}$ & $(q^2)_n^{\vec{P},\,\Lambda}$ & $ME_{FV}$ & $ME_{IV}^{BWI}$ & $ME_{IV}^{BWII}$ \cr
\hline
$\sqrt{ 2 }$ & $B_3$ & $0$          & $(\frac{2 \pi}{L} E_\pi)$ & $0.4603(87)$ & $0.0484(52)$ & $0.0473(73)$ & $6.5(1.1)$ & $6.4(1.1)$ \cr
$\sqrt{ 2 }$ & $B_3$ & $1$          & $(\frac{2 \pi}{L} E_\pi)$ & $0.4603(87)$ & $-0.0440(39)$ & $0.044(14)$ & $6.0(2.0)$ & $5.9(1.9)$ \cr
$\sqrt{ 2 }$ & $B_3$ & $1$          & $(\frac{2 \pi}{L} (E_\pi - E_n))$ & $0.4603(87)$ & $0.0331(39)$ & $0.043(11)$ & $5.9(1.6)$ & $5.8(1.6)$ \cr
$\sqrt{ 2 }$ & $B_3$ & $1$          & $(\frac{2 \pi}{L} E_\pi)$ & $0.4603(87)$ & $0.0331(39)$ & $0.049(16)$ & $6.7(2.2)$ & $6.6(2.2)$ \cr
$\sqrt{ 2 }$ & $B_3$ & $\sqrt{ 2 }$ & $(\frac{2 \pi}{L} (E_\pi - E_n))$ & $0.4603(87)$ & $-0.0360(30)$ & $0.039(20)$ & $5.3(2.7)$ & $5.3(2.7)$ \cr
$\sqrt{ 3 }$ & $A_2$ & $1$          & $(\frac{2 \pi}{L} E_n)$ & $0.4371(98)$ & $0.0035(44)$ & $0.0309(61)$ & $2.50(67)$ & $2.55(69)$ \cr
$\sqrt{ 3 }$ & $A_2$ & $1$          & $(\frac{2 \pi}{L} E_n)$ & $0.4827(89)$ & $0.0257(46)$ & $0.0388(84)$ & $3.8(2.0)$ & $3.8(2.0)$ \cr
$\sqrt{ 3 }$ & $A_2$ & $\sqrt{ 2 }$ & $(\frac{2 \pi}{L} E_n)$ & $0.4371(98)$ & $0.0094(34)$ & $0.039(12)$ & $3.1(1.2)$ & $3.2(1.2)$ \cr
$\sqrt{ 3 }$ & $A_2$ & $\sqrt{ 2 }$ & $(\frac{2 \pi}{L} E_n)$ & $0.4827(89)$ & $0.0268(36)$ & $0.045(16)$ & $4.5(2.6)$ & $4.5(2.6)$ \cr
$\sqrt{ 3 }$ & $E$ & $0$            & $3(\frac{2 \pi}{L} E_\pi)$ & $0.4501(95)$ & $0.0293(58)$ & $0.098(13)$ & $11.6(3.7)$ & $11.8(3.5)$ \cr
$\sqrt{ 3 }$ & $E$ & $0$            & $3(\frac{2 \pi}{L} E_\pi)$ & $0.5178(80)$ & $0.0746(57)$ & $0.098(13)$ & $4.80(68)$ & $4.85(68)$ \cr
$\sqrt{ 3 }$ & $E$ & $1$            & $(\frac{2 \pi}{L} (3E_\pi - E_n))$ & $0.4501(95)$ & $0.0095(44)$ & $0.043(24)$ & $5.1(3.3)$ & $5.2(3.3)$ \cr
$\sqrt{ 3 }$ & $E$ & $1$            & $(\frac{2 \pi}{L} (3E_\pi - 2E_n))$ & $0.4501(95)$ & $0.0095(44)$ & $0.046(20)$ & $5.4(2.9)$ & $5.5(2.9)$ \cr
$\sqrt{ 3 }$ & $E$ & $1$            & $(\frac{2 \pi}{L} E_n)$ & $0.5178(80)$ & $0.0452(45)$ & $0.086(20)$ & $4.2(1.0)$ & $4.2(1.0)$ \cr
$\sqrt{ 3 }$ & $E$ & $1$            & $(\frac{2 \pi}{L} (3E_\pi - E_n))$ & $0.5178(80)$ & $0.0452(45)$ & $0.038(25)$ & $1.9(1.2)$ & $1.9(1.2)$ \cr
$\sqrt{ 3 }$ & $E$ & $\sqrt{ 2 }$   & $3(\frac{2 \pi}{L} (E_\pi - E_n))$ & $0.4501(95)$ & $0.0140(35)$ & $0.119(37)$ & $14.1(6.1)$ & $14.3(5.9)$ \cr
$\sqrt{ 3 }$ & $E$ & $\sqrt{ 2 }$   & $(\frac{2 \pi}{L} E_n)$ & $0.4501(95)$ & $0.0140(35)$ & $0.074(41)$ & $8.7(5.6)$ & $8.9(5.6)$ \cr
$\sqrt{ 3 }$ & $E$ & $\sqrt{ 2 }$   & $(\frac{2 \pi}{L} E_n)$ & $0.5178(80)$ & $0.0425(37)$ & $0.081(43)$ & $4.0(2.1)$ & $4.0(2.2)$ \cr
$\sqrt{ 3 }$ & $E$ & $\sqrt{ 2 }$   & $3(\frac{2 \pi}{L} (E_\pi - E_n))$ & $0.5178(80)$ & $0.0425(37)$ & $0.119(39)$ & $5.8(1.9)$ & $5.9(1.9)$ \cr
\hline
\end{tabular}
\caption{\label{tab:MEpt2}Continuation of Table \protect\ref{tab:MEpt1}.}
\end{table*}

\FloatBarrier

\onecolumngrid

%%%%%%%%%%%%%%%%%%%%%%%%%%%%%%%%%%%%%%%%%%%%%%%%%%%%%%%%%%%%%%%%%%%%%%%%%%%%%%%%%%%%%%%%%%%%%%%%%%%%%%%%%%%%%%%
\section{FIT PARAMETERS}\label{app:MODEL_pars}
%%%%%%%%%%%%%%%%%%%%%%%%%%%%%%%%%%%%%%%%%%%%%%%%%%%%%%%%%%%%%%%%%%%%%%%%%%%%%%%%%%%%%%%%%%%%%%%%%%%%%%%%%%%%%%%
\FloatBarrier

Tables \ref{tab:fitparamsBWI} and \ref{tab:fitparamsBWII} give the fit results for all parametrizations
of the amplitude $\ampV$.

\footnotesize
\begin{table*}[!h]
\begin{center}
\begin{tabular}{|l|c|c|c|c|c|c|c|c|}
\hline
Parameter&       F1 K2     &    F1 K3        &   F2 N1 K2        &      F2 N1 K3   &        F3 N1 M1  &      F3 N1 M2   &      F3 N2 M2      &      F3 N2 M3  \cr
\hline
$A_{00}$ & $ 0.0794(34) $  & $ 0.0799(43) $  &   $ 0.0801(31) $  &  $ 0.0814(34) $  &  $ 0.0834(29) $ &  $ 0.0804(32) $ &   $ 0.0785(42) $   & $ 0.0821(45) $ \cr
$A_{01}$ & $ 0.113(15) $   & $ 0.078(50) $   &   $ 0.116(14) $   &  $ 0.088(33) $   &  $ 0.132(13) $  &  $ 0.113(16) $  &   $ 0.107(19) $    & $ 0.037(53) $  \cr
$A_{02}$ & $ 0.109(46) $   & $ 0.004(158) $  &   $ 0.095(38) $   &  $ 0.004(109) $  &                 &  $ 0.096(38) $  &   $ 0.108(48) $    & $ -0.28(18) $  \cr
$A_{10}$ & $ 0.085(28) $   & $ 0.068(52) $   &   $ 0.081(27) $   &  $ 0.076(31) $   &  $ 0.098(27) $  &  $ 0.086(29) $  &   $ 0.085(32) $    & $ 0.053(37) $  \cr
$A_{11}$ & $ 0.35(19) $    & $ 0.30(33) $    &   $ 0.254(67) $   &  $ 0.21(16) $    &  $ 0.146(50) $  &  $ 0.30(13) $   &   $ 0.42(22) $     & $ 0.77(33) $   \cr
$A_{20}$ & $ 0.12(22) $    & $ 0.24(37) $    &                   &                  &                 &                 &   $ 0.28(36) $     & $ 0.22(42) $   \cr
$A_{12}$ &                 & $ 0.13(96) $    &                   &  $ 0.09(32) $    &                 &  $ -0.09(24) $  &   $ -0.37(55) $    & $ 4.7(2.1) $   \cr
$A_{21}$ &                 & $ 0.4(3.5) $    &                   &                  &                 &                 &   $ -0.5(1.3) $    & $ 7.7(3.6) $   \cr
$A_{22}$ &                 &                 &                   &                  &                 &                 &   $ -0.06(2.06) $  & $ -1.3(9.5) $  \cr
$A_{03}$ &                 & $ 0.28(38) $    &                   &  $ 0.24(27) $    &                 &                 &                    & $ 0.77(44) $   \cr
$A_{13}$ &                 &                 &                   &                  &                 &                 &                    & $ -10.8(4.4) $ \cr
$A_{23}$ &                 &                 &                   &                  &                 &                 &                    & $ -24(21) $    \cr
\hline
\end{tabular}
\end{center}
\caption{\label{tab:fitparamsBWI} Fit results for the amplitude parametrizations based on the \textbf{BW I} Breit-Wigner model.}
\end{table*}

\begin{table*}[!h]
\begin{center}
\begin{tabular}{|l|c|c|c|c|c|c|}
\hline
Parameter&        F1 K2     &       F2 N1 K2   &        F2 N1 K3  &       F3 N1 M2    &      F3 N1 M3  &       F3 N2 M3 \cr
\hline
$A_{00}$ &   $ 0.0803(32) $ & $ 0.0806(30) $   &  $ 0.0821(32) $  &   $ 0.0809(30) $  & $ 0.0827(33) $ & $ 0.0835(39) $ \cr
$A_{01}$ &   $ 0.132(14) $  & $ 0.133(14) $    &  $ 0.097(33) $   &   $ 0.131(15) $   & $ 0.108(37) $  & $ 0.045(53) $  \cr
$A_{02}$ &   $ 0.116(45) $  & $ 0.108(38) $    &  $ -0.01(11) $   &   $ 0.109(38) $   & $ -0.01(11) $  & $ -0.33(18) $  \cr
$A_{10}$ &   $ 0.089(27) $  & $ 0.087(26) $    &  $ 0.077(30) $   &   $ 0.091(28) $   & $ 0.068(33) $  & $ 0.054(36) $  \cr
$A_{11}$ &   $ 0.34(19) $   & $ 0.276(68) $    &  $ 0.21(15) $    &   $ 0.31(12) $    & $ 0.30(20) $   & $ 0.77(32) $   \cr
$A_{12}$ &                  &                  &  $ 0.17(32) $    &   $ -0.08(23) $   & $ 0.74(87) $   & $ 5.1(2.1) $   \cr
$A_{20}$ &   $ 0.07(22) $   &                  &                  &                   &                & $ 0.12(39) $   \cr
$A_{21}$ &                  &                  &                  &                   &                & $ 8.4(3.6) $   \cr
$A_{22}$ &                  &                  &                  &                   &                & $ 0.2(9.3) $   \cr
$A_{03}$ &                  &                  &  $ 0.32(27) $    &                   & $ 0.28(28) $   & $ 0.88(44) $   \cr
$A_{13}$ &                  &                  &                  &                   & $ -1.2(1.7) $  & $ -11.6(4.5) $ \cr
$A_{23}$ &                  &                  &                  &                   &                & $ -28(20) $    \cr
\hline
\end{tabular}
\end{center}
\caption{\label{tab:fitparamsBWII} Fit results for the amplitude parametrizations based on the \textbf{BW II} Breit-Wigner model.}
\end{table*}
\normalsize

\twocolumngrid

\FloatBarrier

\providecommand{\href}[2]{#2}
\begingroup\raggedright

\endgroup

%%%%%%%%%%%%%%%%%%%%%%%%%%%%%%%%%%%%%%%%%%%%%%%%%%%%%%%%%
\end{document}